\documentclass{pj}
\usepackage{amsmath}
\usepackage{amsfonts}
\usepackage{subcaption}
\usepackage{graphicx}
\graphicspath{{./pdf}}

\begin{document}
\setcounter{page}{1}
\pjheader{Vol.\ x, y--z, 2023}

\title[Running head] 
{The Influence of Contrast and Temporal Expansion on the Marching-on-in-Time Contrast Current Density Volume Integral Equation}
\footnote{\it Received date}  \footnote{\hskip-0.12in*\, Corresponding
author:~Petrus~Wilhelmus~Nicolaas~van~Diepen (p.w.n.v.diepen@tue.nl).} 
\footnote{\hskip-0.12in\textsuperscript{1} Department of Electrical Engineering, Eindhoven University of Technology, The Netherlands.} 

\author{Petrus~W.N.~van~Diepen\textsuperscript{*, 1}, Martijn~C.~van~Beurden\textsuperscript{1} and Roeland~J.~Dilz\textsuperscript{1}}

\runningauthor{van~Diepen, van~Beurden and Dilz}

\tocauthor{FistName1~LastName1 and FistName1~LastName1} 

\begin{abstract}
The contrast current density volume integral equation, discretized with piecewise constant spatial basis and test functions and Dirac-delta temporal test functions and the piecewise polynomial temporal basis functions, results in a causal implicit marching-on-in-time scheme that we refer to as the marching-on-in-time contrast current density volume integral equation (MOT-JVIE). The companion matrix stability analysis of the MOT-JVIE solver shows that for a fixed spatial and temporal step size, the stability is independent of the scatterer's dielectric contrast for quadratic spline temporal basis functions. Whereas, Lagrange and cubic spline exhibit instabilities at higher contrast. We relate this stability performance to the expansion and testing procedure in time. We further illustrate the capabilities of the MOT-JVIE based on quadratic spline temporal basis functions by: comparing the MOT-JVIE solution to time-domain results from literature and frequency-domain results from a commercial combined field integral equation solver. Finally, we present a long time sequence for a high-constrast scatterer discretized with 24,000 spatial unknowns.
\end{abstract}


\setlength {\abovedisplayskip} {6pt plus 3.0pt minus 4.0pt}
\setlength {\belowdisplayskip} {6pt plus 3.0pt minus 4.0pt}

\

\section{Introduction}
\label{sc:Introduction}
Time domain Maxwell solvers are more suitable for simulations dealing with non-linearity, wideband electromagnetic excitation and multiphyiscal coupling than their frequency-domain counterparts~\cite{Ergul2019,Jin2019,Zhang2020}. The differential-equation-based time-domain Maxwell solvers, like the finite difference time domain (FDTD) method and the finite element time domain (FETD) method, are often preferred for these types of simulations~\cite{Jin2019,Zhang2020,Sankaran2019}. Another type of time-domain Maxwell solver is the time-domain integral equation (TDIE), which is based on the Green function. TDIEs have two major advantages over their differential-equation-based counterparts~\cite{Ergul2019,Sankaran2019}: 1) the TDIEs do not need a proper truncation of the computational domain, as they satisfy the radiation conditions owing to the use of the Green function; 2) the TDIEs do not need to include the background medium in their computational domain. These advantages reduce the computational domain and could therefore reduce the computational effort of simulations if the TDIEs are applied.

In electromagnetics, the classes of TDIEs can be subdivided into two subclasses, the time-domain surface integral equations (TDSIEs) and the time-domain volume integral equations (TDVIEs)~\cite{DeHoop2008}. The TDVIEs are based on the volume equivalence theorem and use volume equivalent sources to represent wave propagation in penetrable homogeneous and inhomogeneous materials~\cite{DeHoop2008}. Although TDVIEs can be used to represent the electromagnetic interaction with polarizing, magnetizing and lossy propagation media, research over the years has mainly focused on TDVIE solvers dealing with the polarization of dielectrics. The developed TDVIE solvers for wave propagation in dielectrics can be distinguished on multiple aspects. The first aspect is the used unknown quantity. Formulations of the TDVIEs exist with the electric flux density ~\cite{Gres2001,Shanker2004,Kobidze2005,Shi2011bi-iso,Shi2011dispersive,Shi2011Loss,Sayed2015}, the electric field intensity~\cite{Al-Jarro2012,Liu2016}, the contrast current density~\cite{Hu2016,Cao2017,Chen2021}, the electric source vector~\cite{Tong2016}, the magnetic field intensity~\cite{Sayed2020} or a combination of these quantities~\cite{Ulku2015,Sayed2016,Tao2017,Sayed2022}. The second aspect that distinguishes the schemes is the spatial discretization, where the choice is related to the continuity requirements of the unknown quantity used. Implementations exist based on the Schaubert-Wilton-Glisson (SWG) functions~\cite{Gres2001,Shanker2004,Kobidze2005,Sayed2015}, half-SWG functions~\cite{Hu2016}, a combination of half-SWGs and full-SWGs~\cite{Ulku2015,Sayed2016,Sayed2022}, conformal basis functions defined on curvilinear hexahedral
elements~\cite{Shi2011bi-iso,Shi2011dispersive,Shi2011Loss}, pulse basis functions with Dirac-delta test functions~\cite{Liu2016}, fully linear curl-conforming (FLC) basis
functions with FLC or Dirac-delta test functions~\cite{Sayed2020} and Nyström discretizations~\cite{Tong2016,Cao2017,Tao2017,Chen2021}. The third aspect that distinguishes the schemes is the evaluation of the unknown over time, where we can make the distinction between marching-on-in-degree (MOD)~\cite{Shi2011bi-iso,Shi2011dispersive,Shi2011Loss,Tong2016}, implicit causal marching-on-in-time (MOT)~\cite{Gres2001,Shanker2004,Kobidze2005,Cao2017,Tao2017,Sayed2020,Chen2021}, implict non-causal MOT~\cite{Sayed2015}, and explicit MOT~\cite{Al-Jarro2012, Ulku2015, Hu2016,Liu2016,Sayed2016,Sayed2020,Chen2021,Sayed2020,Sayed2022}. The MOD schemes discretize time with a set of orthonormal basis functions and eliminate the time variable from their matrix equations. The solution is obtained by solving a full matrix equation for each of the orthonormal basis functions. Implicit and explicit MOT-schemes both employ Dirac-delta temporal test functions, but expand the temporal part of the unknown differently. The implicit causal MOT-schemes include a temporal expansion of the unknown in basis functions that result in a matrix equation, where the unknowns at each time step do not depend on the unknowns at future time steps. The solution is obtained by solving a sparse matrix equation at each time step. The implicit non-causal MOT-schemes also include a temporal expansion of the unknown in basis functions, however these temporal basis functions extend towards the future. Therefore the resulting matrix equation has unknowns at each time step that do depend on the unknowns at future time steps. The solution is obtained by means of extrapolation in combination with solving a sparse matrix equation at each time step. The explicit MOT-schemes do not include a temporal expansion, but recast the matrix equation into an ordinary differential equation (ODE) in time by employing temporal interpolation. The solution at each time step is obtained using a predictor-corrector scheme for the ODE in combination with solving the Gram matrix equation at each prediction and correction step.

Although all the aforementioned versions of TDVIE solvers are able to compute the electromagnetic scattering due to the presence of a dielectric body, instabilities arise when the dielectric contrast of the scatterer is increased~\cite{Sayed2015}. Only two methods have been shown to be able to deal with high dielectric contrast. One is the implicit non-causal MOT-scheme expanding the electric flux density spatially with SWGs and temporally with approximate prolate spherical wave (APSW) functions~\cite{Sayed2015} and the other is the explicit MOT-scheme expanding the magnetic field intensity using spatially the FLC functions and temporally the fourth-order Lagrange interpolation polynomials~\cite{Sayed2020}. To the best of the authors' knowledge, no implicit causal MOT-scheme based on the TDVIE is documented in literature that remains stable for high dielectric contrast scatterers.

Here, we present an implicit causal MOT-scheme, which we refer to as the marching-on-in-time contrast current density volume integral equation (MOT-JVIE) solver. The MOT-JVIE solver has the contrast current density as unknown. The contrast current density is spatially discretized using piece-wise constant basis and test functions, in line with the frequency-domain scheme presented in~\cite{Polimeridis2014}, which will restrict the spatial part of the contrast current density to the appropriate solution space, i.e. $L^2(\mathbb{R}^3)$~\cite{Beurden2003,Beurden2007,VanBeurden2008}. The temporal discretization consists of Dirac-delta test functions to obtain a MOT-scheme and we show that for a fixed spatial and temporal step size, the stability is independent of the scatterer's dielectric contrast for quadratic spline temporal basis functions. Whereas, other temporal basis functions commonly applied in literature exhibit instabilities at higher contrast. We relate this stability performance to the expansion and testing procedure in time.

This paper is organized as follows. In Section~\ref{sc:MOT-JVIE}, we formulate the TDVIE with the contrast current density as the unknown quantity, discretize this equation and explain how to compute the resulting matrix elements. In Section~\ref{sc:ScatteringSetup}, we introduce the scattering setup that is used in the numerical experiments in subsequent sections. In Section~\ref{sc:Continuity}, we analyze the MOT-JVIE stability behavior for different temporal basis functions. In Section~\ref{sc:Results}, we illustrate the capabilities of the MOT-JVIE based on quadratic spline temporal basis functions by: comparing the MOT-JVIE solution to time-domain results from literature and frequency-domain results from a commercial combined field integral equation solver, and running a long time sequence for a high-contrast dielectric scatterer discretized with 24,000 spatial unknowns. We present our conclusions in Section~\ref{sc:Conclusion}.

\section{MOT-JVIE} \label{sc:MOT-JVIE}
\subsection{Formulation} \label{sc:Formulation}
A dielectric object, occupying a volume $V_\varepsilon$, resides in a homogeneous background medium with permeability $\mu_0$, permittivity $\varepsilon_0$ and resulting wave propagation speed $c_0 = 1/\sqrt{\varepsilon_0 \mu _0}$. The permittivity inside $V_\varepsilon$ is position dependent and is defined as $\varepsilon (\mathbf{r}) = \varepsilon_0\varepsilon_r(\mathbf{r})$, where $\varepsilon_r(\mathbf{r}) \geq 1$ is the relative permittivity. The incident electric field $\mathbf{E}^i(\mathbf{r},t)$, which arrives at the object at $t=0$ or later, will induce a contrast current density, $\mathbf{J}_\varepsilon(\mathbf{r},t)$, in $V_\varepsilon$ after $t=0$. The induced contrast current generates the scattered electric field $\mathbf{E}^s(\mathbf{r},t)$ in accordance with a convolution with the Green function \cite{DeHoop2008}, i.e.
\begin{equation}
     \varepsilon_0\frac{\partial}{\partial t} \mathbf{E}^s (\mathbf{r},t)  =  - \mathbf{J}_\varepsilon(\mathbf{r},t) + \nabla \times \mathbf{H}^s,
\end{equation}
\begin{equation} \label{eq:MagFieldStrength}
    \mathbf{H}^s(\mathbf{r},t) = \nabla \times \iiint_{V_\varepsilon} \frac{\mathbf{J}_\varepsilon(\mathbf{r}',\tau)}{4 \pi R} \mathrm{d}V',
\end{equation}
where $\mathbf{r}$ and $\mathbf{r'}$ are the observer and source spatial coordinates, respectively, $t$ is time, $R=|\mathbf{r}-\mathbf{r}'|$ is distance between source and observer, $\tau = t - \frac{R}{c_0}$ is the retarded time function, $\nabla \times$ is the curl operator with respect to the observer coordinates and $\mathrm{d}V'$ is the infinitesimal volume element over the source coordinates. Together, the incident and scattered electric fields represent the total electric field, $\mathbf{E}(\mathbf{r},t) = \mathbf{E}^i(\mathbf{r},t) + \mathbf{E}^s(\mathbf{r},t)$, which also relates to the contrast current density in the following way~\cite{DeHoop2008},
\begin{equation} \label{eq:ContrastCurrentDensity}
    \mathbf{J}_{\varepsilon}(\mathbf{r},t) = (\varepsilon(\mathbf{r})-\varepsilon_0) \frac{\partial}{\partial t} \mathbf{E}(\mathbf{r},t).
\end{equation}
By combining the above equations, we obtain the time-domain contrast current density volume integral equation (TD-JVIE),
\begin{equation} \label{eq:TDJVIE}
    ( \varepsilon_r(\mathbf{r}) -1)  \varepsilon_0\frac{\partial}{\partial t} \mathbf{E}^i(\mathbf{r},t)  =  \varepsilon_r(\mathbf{r}) \mathbf{J}_\varepsilon(\mathbf{r},t) -  (\varepsilon_r(\mathbf{r})-1) \nabla \times \nabla \times \iiint_{V_\varepsilon} \frac{\mathbf{J}_\varepsilon(\mathbf{r}',\tau)}{4 \pi R} \mathrm{d}V'.
\end{equation}
Equation~\eqref{eq:TDJVIE} illustrates an advantage the TD-JVIE has over the alternatives formulations with electric field intensity and electric flux density as unknown, namely, $\mathbf{J}_\varepsilon(\mathbf{r},t) = 0$ automatically if $\varepsilon(\mathbf{r}) = \varepsilon_0$.

\subsection{Discretization} \label{sc:Discretization} 
To find a numerical approximation for the TD-JVIE~\eqref{eq:TDJVIE} solution, we discretize the equation by expanding the contrast current density as
\begin{equation} \label{eq:J_basis}
    \mathbf{J}_\varepsilon(\mathbf{r},t) = \sum_{\alpha = x,y,z} \sum_{m'=1}^M \sum_{n'=1}^N J^\alpha_{m',n'} \mathbf{f}^{\alpha}_{m'}(\mathbf{r}) T_{n'}(t)
\end{equation}
and introducing the testing operator
\begin{equation}  \label{eq:TestOperator}
    \mathcal{T}^\beta_{m,n}(\mathbf{g}) = \int \delta (t-n\Delta t) \iiint \mathbf{f}^\beta_{m}(\mathbf{r}) \cdot \mathbf{g}(\mathbf{r},t) \mathrm{d}V \mathrm{d}t ,
\end{equation}
for $m = 1,\ldots,M$, $n = 1,\ldots,N$ and $\beta = x, y, z$. In Equation~\eqref{eq:J_basis}, $J_{m',n'}^\alpha$ is an expansion coefficient. In Equation~\eqref{eq:TestOperator}, $\mathbf{g}(\mathbf{r},t)$ is a general three dimensional vector function depending on both space and time, the operator $\cdot$ is the scalar dot product between two three-dimensional vector functions, $\mathrm{d}t$ is the infinitesimal time element, and $\mathrm{d}V$ is the infinitesimal volume element over the observer coordinates. The functions $\mathbf{f}_{m'}^\alpha$~\eqref{eq:J_basis} and $\mathbf{f}_m^\beta$~\eqref{eq:TestOperator} are the spatial basis and test function, respectively. To restrict the solution for $\mathbf{J}_\varepsilon$ to the appropriate spatial function space, i.e. $L^2(\mathrm{I\!R}^3)$~\cite{Beurden2003,Beurden2007,VanBeurden2008}, we use piece-wise constant functions as basis and test functions~\cite{Polimeridis2014}. The $m'$-th basis vector function is then defined for all three Cartesian unit vectors, i.e. $\hat{\boldsymbol{\alpha}} = \hat{\mathbf{x}}, \hat{\mathbf{y}}, \hat{\mathbf{z}}$, as
\begin{equation} \label{eq:SpatialBasis}
    \mathbf{f}^\alpha_{m'} (\mathbf{r}) = \begin{cases}
    \hat{\boldsymbol{\alpha}}, & \text{for} \, \mathbf{r} \in V_{m'}, \\
    \mathbf{0}, & \text{for} \, \mathbf{r}  \notin V_{m'}, \\
    \end{cases}
\end{equation}
where $V_{m'}$ is the volume occupied by a voxel of dimension $\Delta x \times \Delta y \times \Delta z$ centered around $\mathbf{r}_{m'}$. In a similar way we define the $m$-th test vector function $\mathbf{f}_m^\beta$ defined on the voxel centered around $\mathbf{r}_m$ occupying the volume $V_m$, where $\mathbf{f}_{m}^\beta = \mathbf{f}_{m'}^\alpha$ if $\beta = \alpha$ and $m = m'$. All $M$ voxels are positioned in a regularized voxel grid, where the center of any two voxels are separated by an integer times $\Delta x$, $\Delta y$ and/or $\Delta z$ in the accompanying Cartesian directions. The function $\delta(t-n \Delta t)$ and $T_{n'}(t)$ are the $n$-th temporal test and $n'$-th temporal basis function, respectively. The test functions $\delta(t-n \Delta t)$ are the Dirac-delta functions. The use of Dirac-delta functions prevents an expensive evaluation of the time integral in~\eqref{eq:TestOperator} and in combination with a properly chosen temporal basis function the discretized equation can be solved as a marching-on-in-time (MOT)-scheme, i.e. unknowns at each time step do not depend on the unknowns at future time steps~\cite{Poggio1973}. The numerical stability of a MOT-scheme will depend on the choice of temporal basis function, $T_{n'}$,~\cite{Sayed2015}. In Section~\ref{sc:Continuity} we will demonstrate that for several temporal basis functions the MOT-JVIE is or becomes unstable when we increase the scatterer's dielectric contrast and that the quadratic spline temporal basis function is a basis function for which the numerical stability does not depend on the electric contrast of the scatterer. The quadratic spline temporal basis functions are defined as
\begin{equation} \label{eq:QuadSpline}
    T_{n'}(t) = \begin{cases}
    \frac{1}{2}\left(\frac{t-n'\Delta t}{\Delta t}\right)^2 + \left(\frac{t-n'\Delta t}{\Delta t}\right) + \frac{1}{2},& t \in I_1, \\
    -\left(\frac{t-n'\Delta t}{\Delta t}\right)^2 + \left(\frac{t-n'\Delta t}{\Delta t}\right) + \frac{1}{2},& t \in I_2, \\
    \frac{1}{2}\left(\frac{t-n'\Delta t}{\Delta t}\right)^2 -2 \left(\frac{t-n'\Delta t}{\Delta t}\right) + 2,& t \in I_3, \\
    0,& \text{else},
    \end{cases}
\end{equation}
where $I_1 = ((n'-1)\Delta t, n'\Delta t]$, $I_2 = (n'\Delta t, (n'+1)\Delta t]$ and $I_3 = ((n+1) \Delta t, (n'+2)\Delta t]$. By default, the MOT-JVIE is based on quadratic spline basis functions, unless stated otherwise.

By substituting~\eqref{eq:J_basis} in~\eqref{eq:TDJVIE} and applying the testing operator~\eqref{eq:TestOperator}, we arrive at a block-matrix equation of the form
\begin{equation} \label{eq:MatrixEquation}
    \begin{bmatrix}
    \mathbf{Z}_0 & \\
    \mathbf{Z}_1 & \mathbf{Z}_0 \\
    \vdots & \ddots & \ddots \\
    \mathbf{Z}_\ell & \cdots & \mathbf{Z}_1 &  \mathbf{Z}_0 \\
     & \ddots & & \ddots & \ddots\\
    & & \mathbf{Z}_\ell & \cdots & \mathbf{Z}_1 &  \mathbf{Z}_0
    \end{bmatrix} 
    \begin{bmatrix}
    \mathbf{J}_1 \\
    \mathbf{J}_2 \\
    \vdots \\
    \vdots \\
    \vdots \\
    \mathbf{J}_N
    \end{bmatrix} = \begin{bmatrix}
    \mathbf{E}_1 \\
    \mathbf{E}_2 \\
    \vdots \\
    \vdots \\
    \vdots \\
    \mathbf{E}_N
    \end{bmatrix},
\end{equation}
where the blocks of the unknown vector are defined as
\begin{equation} \label{eq:CurrentVector}
    \mathbf{J}_n = \left[J_{1,n}^x,J_{1,n}^y,J_{1,n}^z,\ldots,J_{M,n}^x,J_{M,n}^y,J_{M,n}^z\right]^T,
\end{equation}
the blocks of the excitation vector are defined as
\begin{equation} \label{eq:ExcitationVector}
    \mathbf{E}_n = \left[E_{1,n}^x,E_{1,n}^y,E_{1,n}^z,\ldots,E_{M,n}^x,E_{M,n}^y,E_{M,n}^z\right]^T,
\end{equation}
with 
\begin{equation} \label{eq:EsingleV}
    E^\beta_{m,n} = \iiint_{V_m} \hat{\boldsymbol{\beta}} \cdot ( \varepsilon_r(\mathbf{r})-1)  \varepsilon_0\frac{\partial}{\partial t} \mathbf{E}^i(\mathbf{r},n \Delta t) \mathrm{d}V,
\end{equation}
where $\hat{\boldsymbol{\beta}}$ represents one of the three Cartesian unit vectors $\hat{\mathbf{x}}$, $\hat{\mathbf{y}}$, or $\hat{\mathbf{z}}$,
and the blocks in the matrix, i.e. $\mathbf{Z}_{n-n'}$, are called the interaction matrices, which have to be precomputed for $n-n' = 0$ up to 
\begin{equation} \label{eq:Ell}
    \ell = \left \lfloor{\frac{R\textsubscript{max}}{c \Delta t}} \right \rfloor + p,
\end{equation}
where $R\textsubscript{max}$ is the maximum distance between source and observation point in the computational domain, $\lfloor \cdot \rfloor$ is the floor operator and $p$ is the order of the polynomial used for the temporal basis function $T_{n'}(t)$, i.e. $p = 2$ when employing~\eqref{eq:QuadSpline}. The elements of the interaction matrices are discussed in Section~\ref{sc:MatrixElements}.

As the matrix equation in~\eqref{eq:MatrixEquation} is a block-Toeplitz lower-block-triangular matrix, forward block substitution is used to find the unknown expansion coefficients $\mathbf{J}_n$~\eqref{eq:CurrentVector} at each $n$. Subsequently, we define the marching-on-in-time contrast current density volume integral equation (MOT-JVIE)
\begin{equation} \label{eq:MOT-JVIE}
    \mathbf{Z}_0 \mathbf{J}_n = \mathbf{E}_{n} - \sum_{n'=n-\ell}^{n-1} \mathbf{Z}_{n-n'} \mathbf{J}_{n'}.
\end{equation}
Solving Equation~\eqref{eq:MOT-JVIE} requires solving $\mathbf{Z}_0 \mathbf{J}_n = \mathbf{V}$ at every time step. The MOT-JVIE is classified as an implicit causal MOT-scheme, as we expand the temporal part of the contrast current density with temporal basis functions and the unknowns at time $n$ do not depend on the unknowns at future time steps.

\subsection{Computing the interaction matrix elements} \label{sc:MatrixElements}
Before we can employ the MOT-JVIE~\eqref{eq:MOT-JVIE} to compute the unknown contrast current density, we have to precompute the interaction matrices $\mathbf{Z}_{n-n'}$ for $n-n' = 0, \ldots, \ell$~\eqref{eq:Ell}. Given the structure of the current density vector $\mathbf{J}_n$ in~\eqref{eq:CurrentVector} and the excitation vector $\mathbf{E}_n$ in~\eqref{eq:ExcitationVector}, the interaction matrices $\mathbf{Z}_{n-n'}$ in~\eqref{eq:MOT-JVIE} will consist of $m \times m'$ block matrices $\mathbf{Z}_{m,m',n-n'}$ of size $3 \times 3$. Each block matrix $\mathbf{Z}_{m,m',n-n'}$ is related to the interaction between two voxels $m$ and $m'$ and the indices of the blocks are related to three Cartesian unit vectors defined on the voxels.  By approximating the relative permittivity function as constant throughout the voxel $m$, i.e. $\varepsilon_m = \varepsilon_r(\mathbf{r}_m)$, the block at block-row $m$ and block-column $m'$ is defined as
\begin{equation} \label{eq:Zblocks}
    \mathbf{Z}_{m,m',n-n'} = \varepsilon_m v \delta_{m,m'} T_{n'}(n \Delta t)\mathbf{I}_3  - (\varepsilon_m-1)\mathbf{C}_{m,m',n-n'},
\end{equation}
The first term on the right-hand-side of~\eqref{eq:Zblocks} is the result of equal Cartesian unit vectors in the spatial basis, $\mathbf{f}_{m'}^\alpha$~\eqref{eq:SpatialBasis}, and test function, $\mathbf{f}_m^\beta$, for overlapping voxels, i.e. $\mathbf{f}_{m'}^\alpha = \mathbf{f}_{m'}^\beta$ at voxel $m = m'$. In that first term $v = \Delta x \Delta y \Delta z$ is the volume of a voxel, $\delta_{m,m'}$ is the Kronecker delta, i.e. $\delta_{m,m'} = 1$ if $m = m'$ and $\delta_{m,m'} = 0$ if $m \neq m'$, and $\mathbf{I}_3$ is the identity matrix of dimension three. The elements of the $3 \times 3$ block matrix $\mathbf{C}_{m,m',n-n'}$ in the second term on the right-hand-side of~\eqref{eq:Zblocks} are defined as
\begin{equation} \label{eq:ZTestVolume}
    {C}^{\beta,\alpha}_{m,m',n-n'} = \iiint \mathbf{f}_{m}^\beta(\mathbf{r}) \cdot \nabla \times  \mathbf{H}^\alpha_{m',n-n'}(\mathbf{r}) \mathrm{d}V,
\end{equation}
where
\begin{equation} \label{eq:ZBasisVolume}
    \mathbf{H}^{\alpha}_{m',n-n'}(\mathbf{r}) = \nabla \times \iiint \frac{\mathbf{f}_{m'}^\alpha(\mathbf{r}') T_{n'}(\tau_n)}{4\pi R} \mathrm{d}V',
\end{equation}
and $\tau_n = \tau(n \Delta t)$. By employing the divergence theorem, we can rewrite the above volume integrals over $V_m$ and $V_{m'}$ to their respective enclosing surfaces $\partial V_m$ and $\partial V_{m'}$
\begin{equation} \label{eq:SurfaceTest}
    C^{\beta,\alpha}_{m,m',n-n'} =  -\iint_{\partial V_m} \left( \hat{\boldsymbol{\beta}} \times \hat{\mathbf{n}}_{m} \right) \cdot \mathbf{H}^{\alpha}_{m',n-n'}(\mathbf{r}) \mathrm{d} A,
\end{equation}
\begin{equation} \label{eq:SurfaceBasis}
    \mathbf{H}^{\alpha}_{m',n-n'}(\mathbf{r}) = \iint_{\partial V_{m'}} \left( \hat{\boldsymbol{\alpha}} \times \hat{\mathbf{n}}_{m'} \right) \frac{T_{n'}(\tau_n)}{4 \pi R} \mathrm{d}A',
\end{equation}
where $\hat{\mathbf{n}}_{m}$ and $\hat{\mathbf{n}}_{m'}$ are the normals on the surfaces $\partial V_{m}$ and $\partial V_{m'}$, respectively, and $\mathrm{d}A$ and $\mathrm{d}A'$ are the infinitesimal surface elements over the observer and source coordinates, respectively. As the temporal basis function, $T_{n'}$, is a polynomial, see~\eqref{eq:QuadSpline}, we end up in Equation~\eqref{eq:SurfaceBasis} with integrals of the form
\begin{equation}
    \iint \frac{1}{R^k} \mathrm{d}A' \text{ for } k = -1,0,1 \, \text{and} \,2.
\end{equation}
Analytic expressions for these integrals have been published in~\cite{Shanker2009,Wout2013} and these are used to compute the value for $\mathbf{H}^{\alpha}_{m',n-n'}(\mathbf{r})$ for any given values of $\alpha$, $m'$, $n$, $n'$ and $\mathbf{r}$. The implementation in~\cite{Wout2013} requires the tolerance values $\epsilon\textsubscript{edge}$ and $\epsilon\textsubscript{vertex}$, which we both set to $10^{-8}$ as recommended in~\cite{Wout2013}. Unfortunately, an analytic expression for the surface integration in~\eqref{eq:SurfaceTest} is not available. However, we can approximate the surface integral numerically using $q$th-order Gauss-Legendre quadrature along each Cartesian direction. The necessary weights and nodes for Gauss-Legendre quadrature are computed in accordance to the Golub-Welsch algorithm~\cite{Golub1969}. Note that, due to numerical approximation of the surface integral in Equation~\eqref{eq:SurfaceTest}, we recommend $c \Delta t \geq \max(\Delta x, \Delta y, \Delta z)$, as smaller time steps will result in more rapid variations of the integrand in the surface integration domain~\eqref{eq:SurfaceTest} which will diminish the accuracy of the numerical approximation of Equation~\eqref{eq:SurfaceTest}. Meeting this $\Delta t$ requirement, we observe in numerical experiments a $q^{-4}$ proportionality of the error in the slowest converging matrix element, i.e. the matrix elements $m = m'$ for which the integrand is singular. Therefore, we choose $q = 5$, as this is accurate enough for our purpose here, i.e. if $\Delta t = \Delta x = \Delta y = \Delta z$ the relative error at the $m = m'$ matrix elements is $0.002$. By combining the analytic evaluation of the integral in~\eqref{eq:SurfaceBasis} with the numerical evaluation of the integral in~\eqref{eq:SurfaceTest}, we can evaluate $\mathbf{Z}_{m,m',n-n'}$ for all $m$, $m'$ and $n-n'$. In practice, it is unnecessary to compute these blocks for all $m$ and $m'$, as uniform meshing with voxels of the same size, i.e. $\Delta x \times \Delta y \times \Delta z$ is fixed, allows for reuse of already computed elements, which reduces computation time.

The remaining volume integral in~\eqref{eq:EsingleV} can also be evaluated using $q$th-order Gauss-Legendre quadrature in all three Cartesian directions. For ease of implementation, we choose the same order as used earlier, i.e. $q = 5$.

\section{Scattering setup} \label{sc:ScatteringSetup}
\begin{figure}[t]
\centerline{\includegraphics[width=0.5\columnwidth,draft=false]{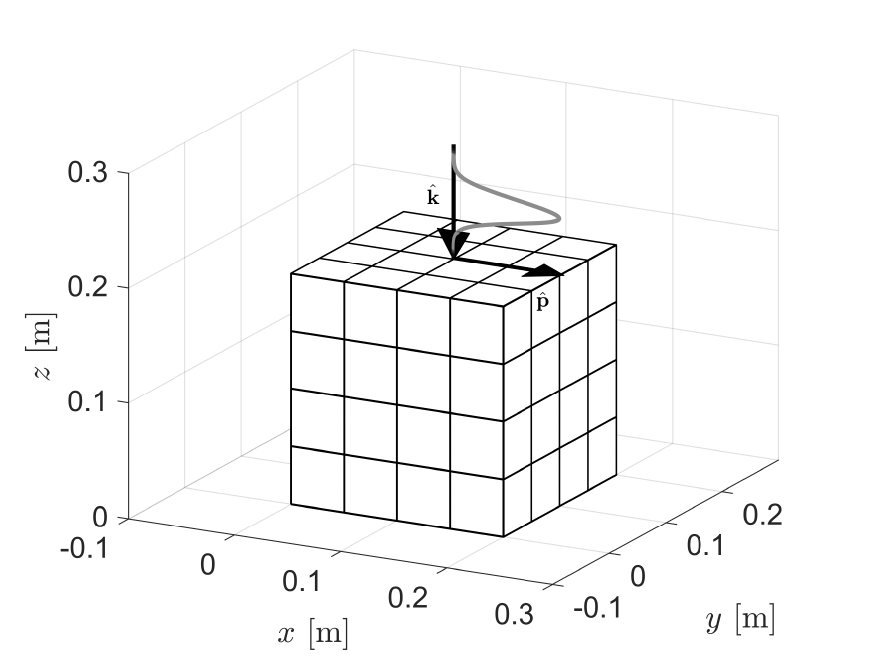}}
\caption{The scattering setup consisting of a discretized dielectric cube with $K = 4$ and the incident Gaussian plane wave, represented by the gray pulse, with polarization $\hat{\mathbf{p}}$ and propagation direction $\hat{\mathbf{k}}$.}
\label{fig:ScatterSetup}
\end{figure}

Through multiple numerical experiments we demonstrate the MOT-JVIE (in)stability for different temporal basis functions in Section~\ref{sc:Continuity} and further illustrate the MOT-JVIE solver capabilities in Section~\ref{sc:Results}. The numerical experiments in both sections are performed on the same scattering setup as shown in Figure~\ref{fig:ScatterSetup}, but for various relative permittivity values. The setup consists of a dielectric cube excited by an electromagnetic excitation. The cube has an edge length of $0.2~\mathrm{m}$ with a center positioned at $\mathbf{r} = (0.1,0.1,0.1)$. The cube is uniformly discretized in all Cartesian directions and $K$ indicates the number of voxels per Cartesian direction. The electromagnetic excitation in all experiments is an $\hat{\mathbf{x}}$-polarized Gaussian plane wave travelling in the negative $\hat{\mathbf{z}}$-direction, i.e.
\begin{equation} \label{eq:GaussPlaneWave}
    \mathbf{E}^i (\mathbf{r},t) = E_0 \frac{4}{w\sqrt{\pi}} \hat{\mathbf{x}} \exp{\left(-\frac{16}{w^2}\left((t-t_0) + \mathbf{r}\cdot \hat{\mathbf{z}})\right)^2 \right)},
\end{equation}
where $E_0$ is the amplitude scaling set to $E_0 = 1~\mathrm{V}/\mathrm{m}$, $w$ is the pulse width in $\mathrm{lm}$ and $t_0$ is the separation time at time $t = 0$ between the Gaussian pulse center and the coordinate system origin in $\mathrm{lm}$. The unit $\mathrm{lm}$ in this context is known as lightmeter, i.e. the time it takes for the wave to travel a distance of $1~\mathrm{m}$.

\section{MOT-JVIE stability analysis} \label{sc:Continuity}
To study the stability performance of multiple MOT-JVIE implementations based on different temporal basis functions, when increasing the scatterer's dielectric contrast, we perform the companion-matrix stability analysis~\cite{Dodson1998} for the dielectric cube discussed in Section~\ref{sc:ScatteringSetup}. The cube is discretized with $K = 6$ and $\Delta t = 0.2/K~\mathrm{lm}$. The eigenvalues $\lambda$ of the MOT-JVIE companion matrices for $\varepsilon_r = 3.2$ and $\varepsilon_r = 100$ are compared in the Sections~\ref{sc:Lagrange} and~\ref{sc:Spline}. In Section~\ref{sc:EigDiscussion} we discuss the implications of the numerical experiment.

\subsection{Numerical results Lagrange temporal basis functions} \label{sc:Lagrange}
\begin{figure}[t]
    \centering \begin{subfigure}{0.4\textwidth}
        \centerline{\includegraphics[width=\textwidth,draft=false]{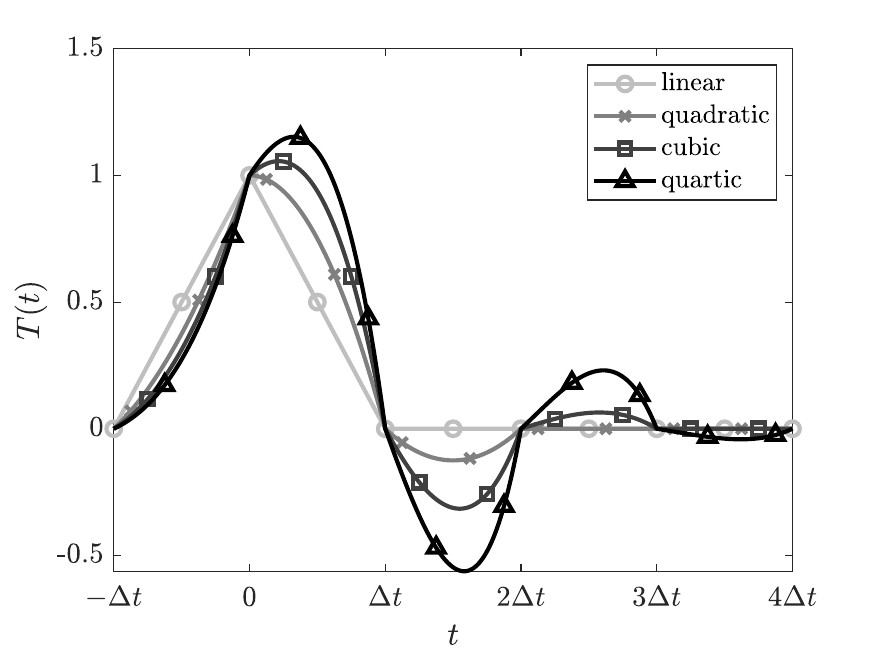}}
        \caption{}
        \label{fig:Lagrange}
    \end{subfigure}
    \begin{subfigure}{0.4\textwidth}
        \centering     
        \includegraphics[width=\textwidth,draft=false]{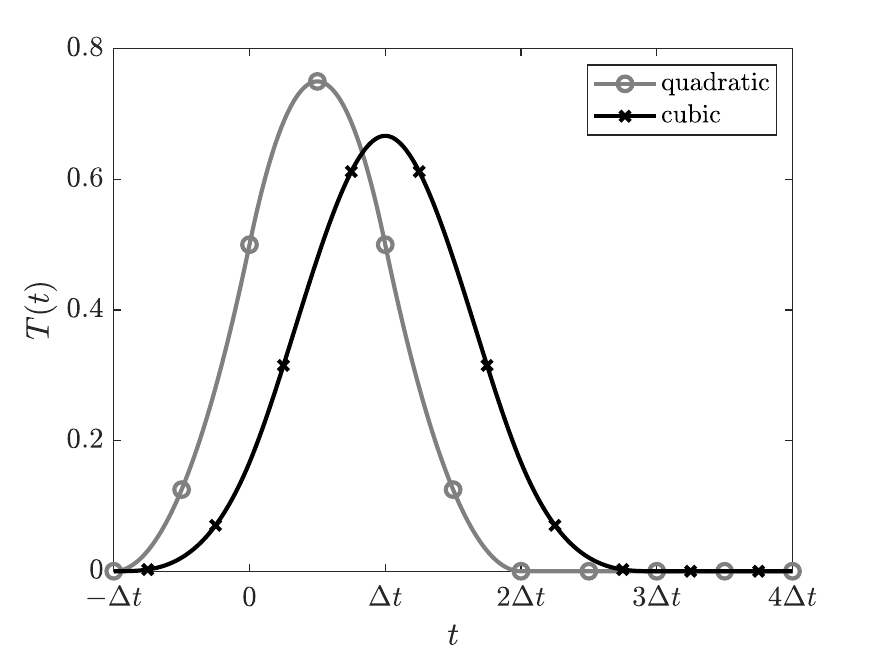}
    \caption{}
    \label{fig:Spline}
    \end{subfigure}
    \caption{(a) The linear, quadratic, cubic and quartic Lagrange temporal basis functions with $\mathcal{C}^0$-continuity. (b) The quadratic and cubic spline temporal basis functions with $\mathcal{C}^1$ and $\mathcal{C}^2$-continuity, respectively.}
    \label{fig:TempBasisFunctions}
\end{figure}

\begin{figure}[p]
\centering
\begin{subfigure}[b]{0.35\textwidth}
    \centerline{\includegraphics[width=\textwidth,draft=false]{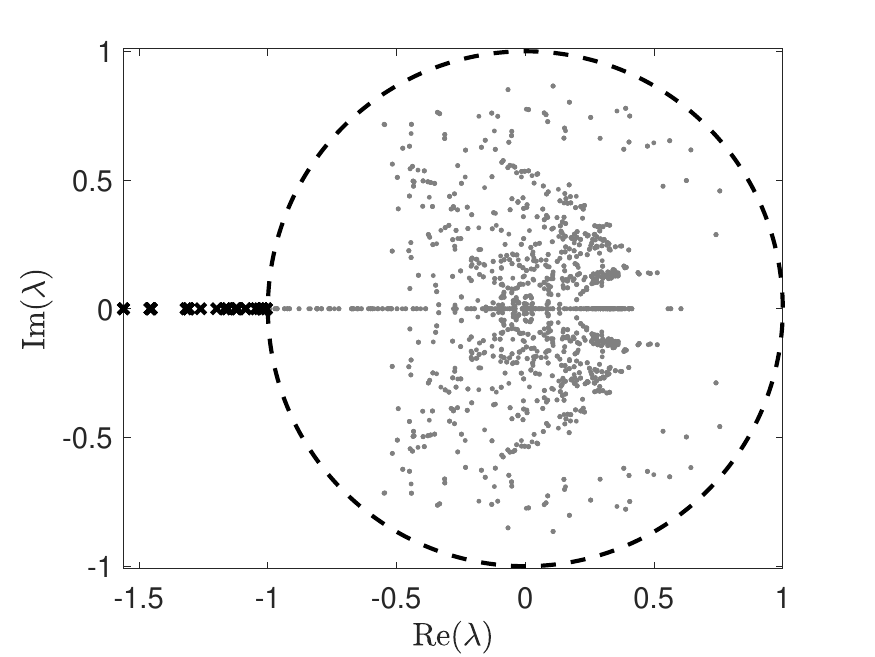}}
    \caption{$1\textsuperscript{st}$-order, $\varepsilon_r = 3.2$}
    \label{fig:EigLin_epsr3_2}
\end{subfigure}
\begin{subfigure}[b]{0.35\textwidth}
    \centerline{\includegraphics[width=\textwidth,draft=false]{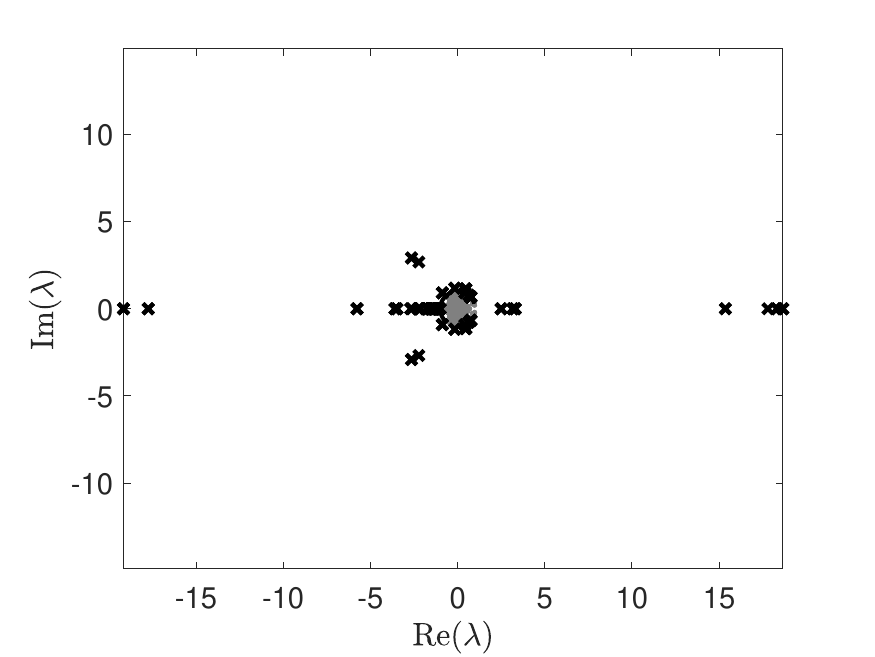}}
    \caption{$1\textsuperscript{st}$-order, $\varepsilon_r = 100$}
    \label{fig:EigLin_epsr100}
\end{subfigure}
\\
\begin{subfigure}[b]{0.35\textwidth}
    \centerline{\includegraphics[width=\textwidth,draft=false]{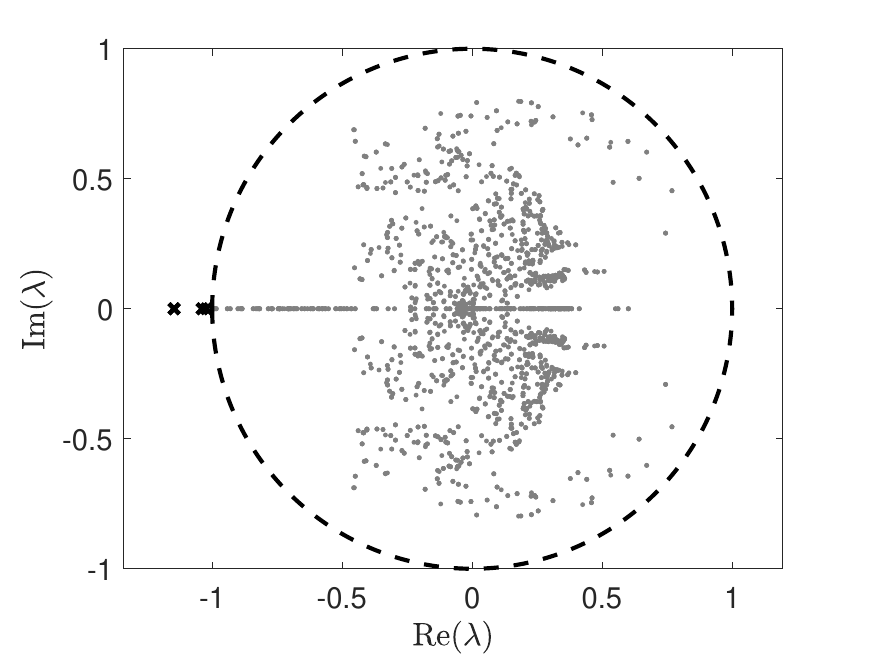}}
    \caption{$2\textsuperscript{nd}$-order, $\varepsilon_r = 3.2$}
    \label{fig:EigQuadLag_epsr3_2}
\end{subfigure}
\begin{subfigure}[b]{0.35\textwidth}
    \centerline{\includegraphics[width=\textwidth,draft=false]{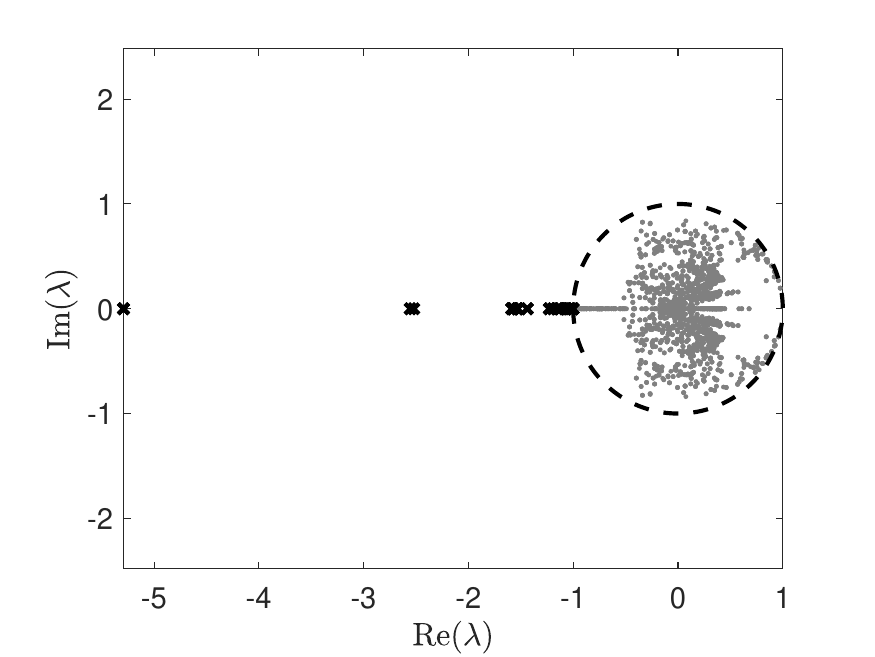}}
    \caption{$2\textsuperscript{nd}$-order, $\varepsilon_r = 100$}
    \label{fig:EigQuadLag_epsr100}
\end{subfigure}
\\
\begin{subfigure}[b]{0.35\textwidth}
    \centerline{\includegraphics[width=\textwidth,draft=false]{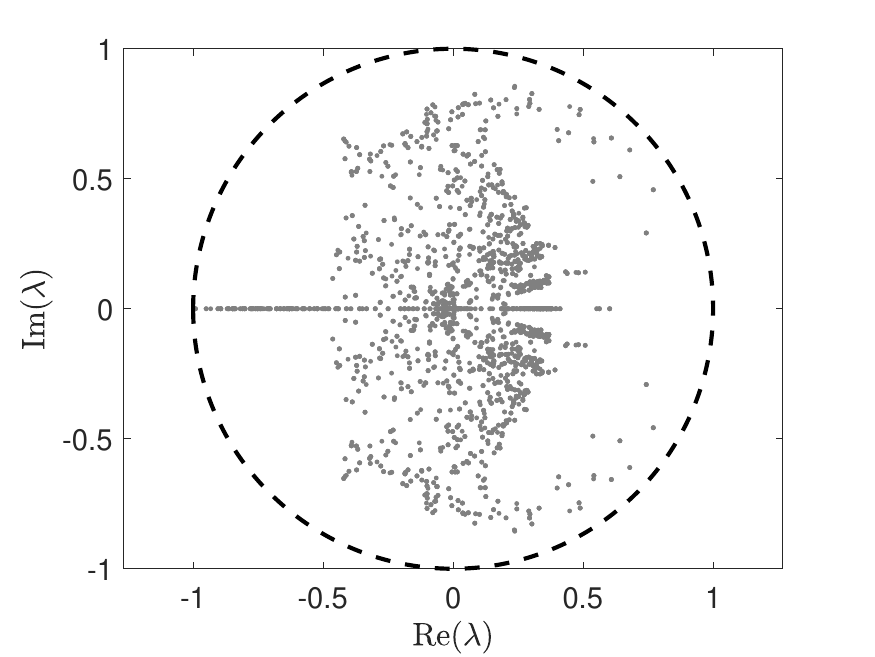}}
    \caption{$3\textsuperscript{rd}$-order, $\varepsilon_r = 3.2$}
    \label{fig:EigCubicLag_epsr3_2}
\end{subfigure}
\begin{subfigure}[b]{0.35\textwidth}
    \centerline{\includegraphics[width=\textwidth,draft=false]{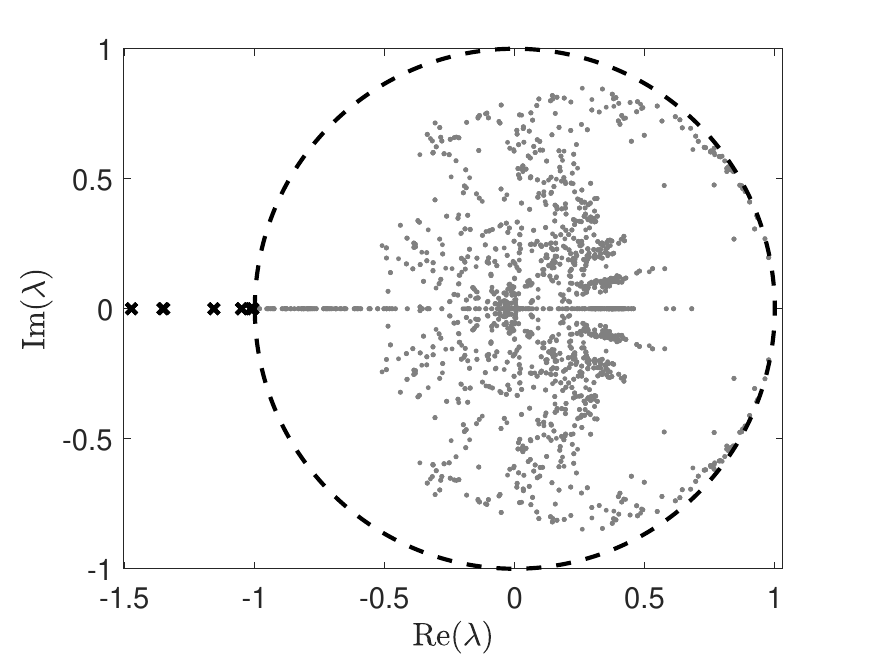}}
    \caption{$3\textsuperscript{rd}$-order, $\varepsilon_r = 100$}
    \label{fig:EigCubicLag_epsr100}
\end{subfigure}
\\
\begin{subfigure}[b]{0.35\textwidth}
    \centerline{\includegraphics[width=\textwidth,draft=false]{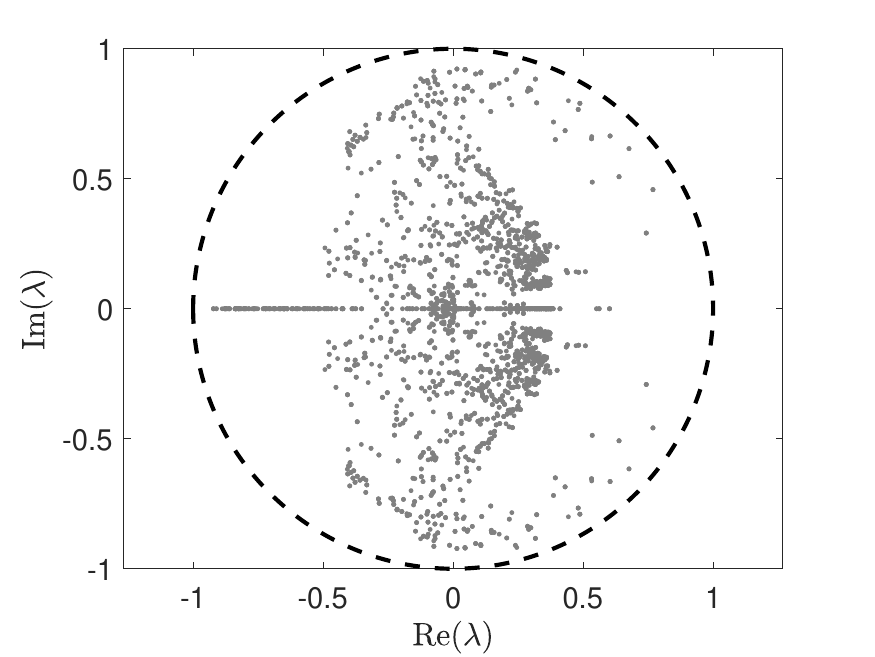}}
    \caption{$4\textsuperscript{th}$-order, $\varepsilon_r = 3.2$}
    \label{fig:EigQuarticLag_epsr3_2}
\end{subfigure}
\begin{subfigure}[b]{0.35\textwidth}
    \centerline{\includegraphics[width=\textwidth,draft=false]{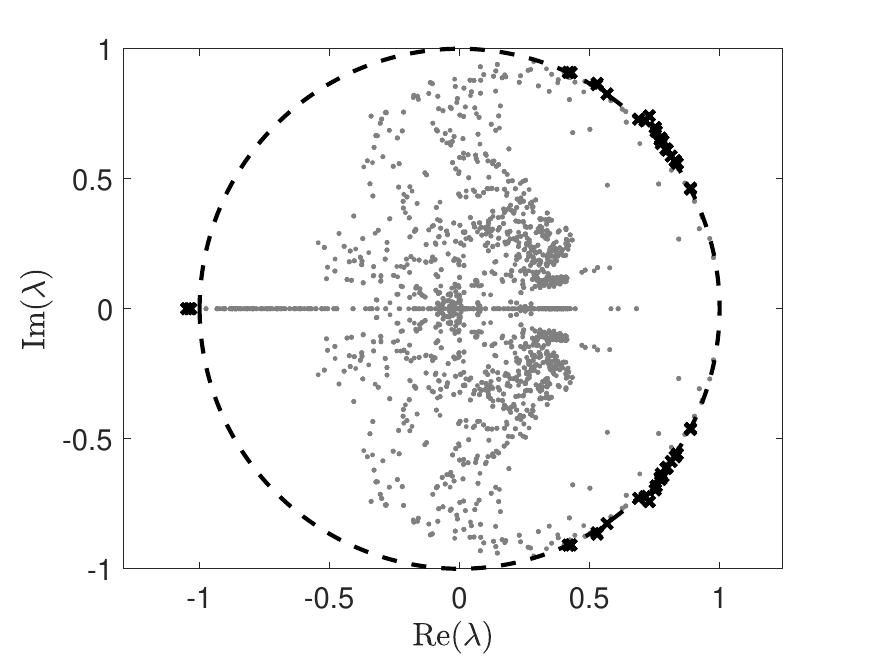}}
    \caption{$4\textsuperscript{th}$-order, $\varepsilon_r = 100$}
    \label{fig:EigQuarticLag_epsr100}
\end{subfigure}
\caption{The MOT-JVIE companion-matrix eigenvalue, $\lambda$, spectrum. The gray $(\cdot)$-markers indicate eigenvalues with $|\lambda|<1$ and the black $(\times)$-markers indicate eigenvalues with $|\lambda| \geq 1$. The companion matrix is constructed with interaction matrices that correspond to the dielectric cube discretized with $K = 6$ and $\Delta t = 0.2/K~\mathrm{lm}$ and different $\mathcal{C}^0$-continuous temporal basis functions and relative permittivities $\varepsilon_r$. The linear temporal basis function with $\varepsilon_r = 3.2$ in (a) and $\varepsilon_r = 100$ in (b). The quadratic Lagrange temporal basis function with $\varepsilon_r = 3.2$ in (c) and $\varepsilon_r = 100$ in (d). The cubic Lagrange temporal basis function with $\varepsilon_r = 3.2$ in (e) and $\varepsilon_r = 100$ in (f). The quartic Lagrange temporal basis function with $\varepsilon_r = 3.2$ in (g) and $\varepsilon_r = 100$ in (h). }
\label{fig:EigLag}
\end{figure}

The implicit causal MOT-schemes found in literature are based on linear~\cite{Cao2017}, quadratic~\cite{Kobidze2005}, cubic~\cite{Gres2001,Chen2021} and quartic~\cite{Shanker2004,Tao2017,Sayed2020} Lagrange temporal basis functions, see Figure~\ref{fig:Lagrange}. If the authors explain their choice, it is related to the number of time-derivatives in the TDVIE they choose to discretize and a higher order than strictly necessary is used to have faster temporal convergence. The TDJVIE~\eqref{eq:TDJVIE} does not contain any direct time derivative, so the equation can be evaluated with all four Lagrange temporal basis functions. The interaction matrices are computed as explained in Section~\ref{sc:MatrixElements} for each of the temporal basis functions and the corresponding companion-matrix eigenvalue spectra are shown in Figure~\ref{fig:EigLag}, where the unit circle to distinguish stability from instability is indicated by the dashed circle. Although the MOT-JVIE based on cubic and quartric Lagrange temporal basis functions are stable for $\varepsilon_r = 3.2$, all MOT-JVIE expressions based on Lagrange temporal basis functions become unstable when the relative permittivity of the scatterer is increased to $\varepsilon_r = 100$. The eigenvalues on the negative real axis are the first to become unstable when increasing the scatterer permittivity, as was also observed for the marching-on-in-time electric flux density volume integral equation based on cubic Lagrange temporal basis functions~\cite{Sayed2015}. Although not included here, reducing the relative permittivity of the scatterer to $\varepsilon_r = 2$ or lower results in a stable MOT-JVIE for all four Lagrange temporal basis functions. 

\subsection{Numerical results Spline temporal basis functions} \label{sc:Spline}
\begin{figure}[t]
\centering
\begin{subfigure}[b]{0.35\textwidth}
    \centerline{\includegraphics[width=\textwidth,draft=false]{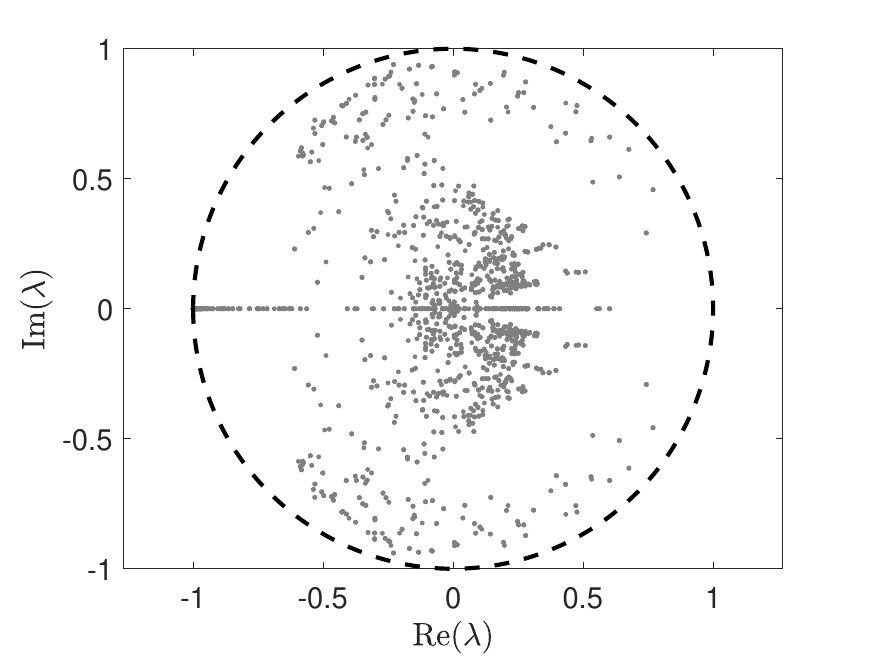}}
    \caption{$2\textsuperscript{nd}$-order, $\varepsilon_r = 3.2$}
    \label{fig:EigQuad_epsr3_2}
\end{subfigure}
\begin{subfigure}[b]{0.35\textwidth}
    \centerline{\includegraphics[width=\textwidth,draft=false]{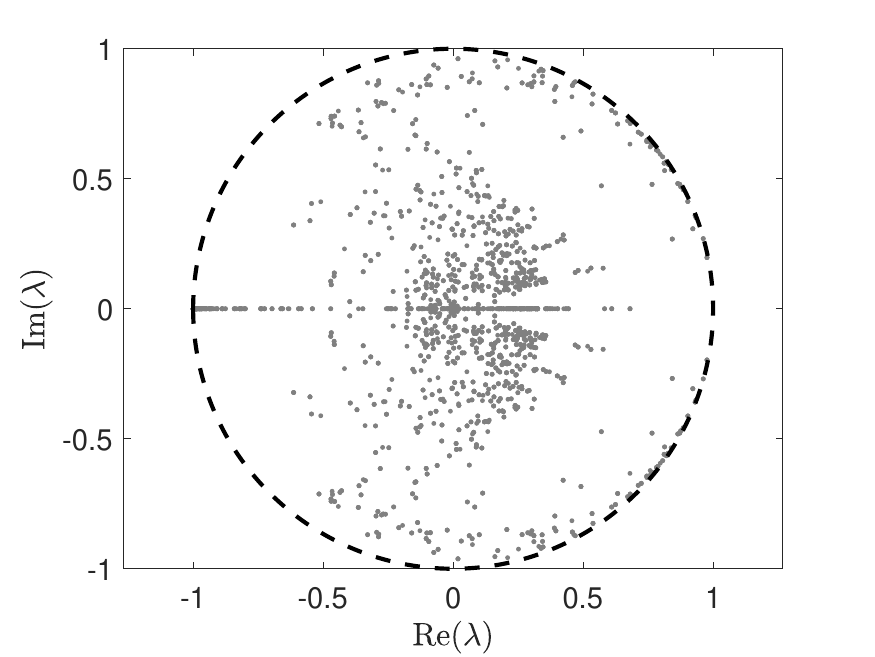}}
    \caption{$2\textsuperscript{nd}$-order, $\varepsilon_r = 100$}
    \label{fig:EigQuad_epsr100}
\end{subfigure}
\\
\begin{subfigure}[b]{0.35\textwidth}
    \centerline{\includegraphics[width=\textwidth,draft=false]{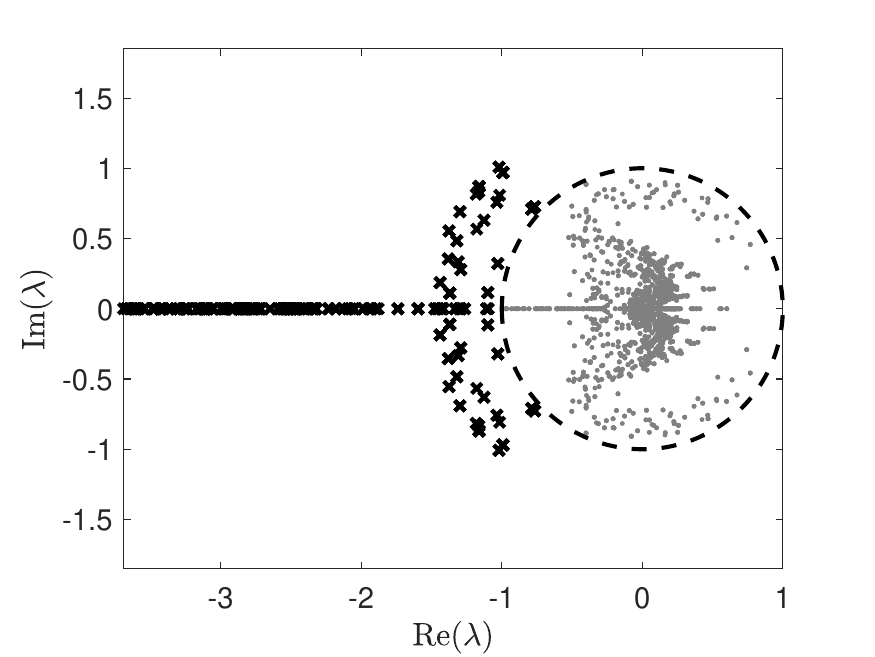}}
    \caption{$3\textsuperscript{rd}$-order, $\varepsilon_r = 3.2$}
    \label{fig:EigCubic_epsr3_2}
\end{subfigure}
\begin{subfigure}[b]{0.35\textwidth}
    \centerline{\includegraphics[width=\textwidth,draft=false]{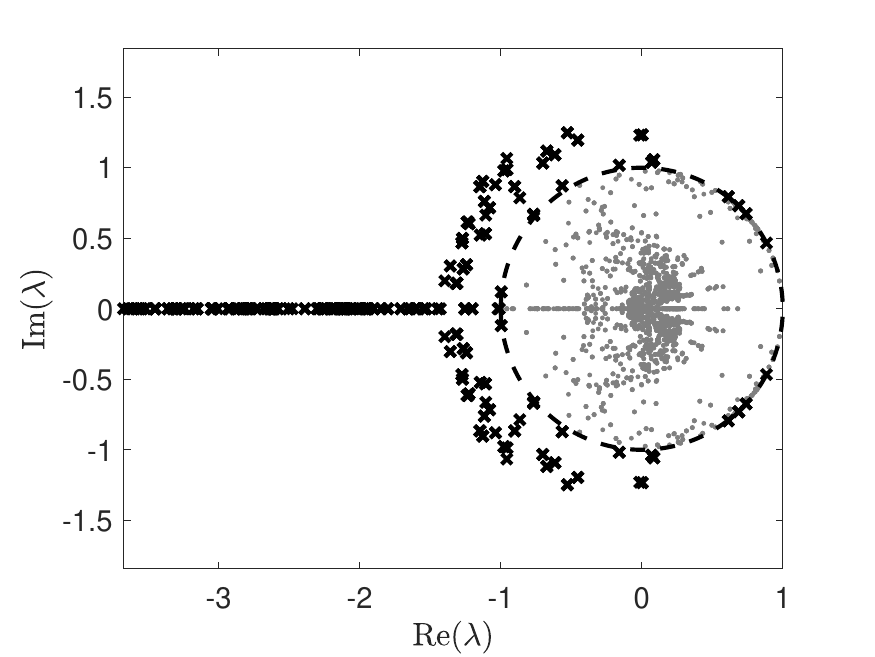}}
    \caption{$3\textsuperscript{rd}$-order, $\varepsilon_r = 100$}
    \label{fig:EigCubic_epsr100}
\end{subfigure}
\caption{The MOT-JVIE companion-matrix eigenvalue spectrum, $\lambda$, for two relative permittivities, i.e. $\varepsilon_r = 3.2$ and $\varepsilon_r = 100$. The gray $(\cdot)$-markers indicate eigenvalues with $|\lambda|<1$ and the black $(\times)$-markers indicate eigenvalues with $|\lambda| \geq 1$. The companion matrix is constructed with interaction matrices that correspond to the dielectric cube discretized with $K = 6$ and $\Delta t = 0.2/K~\mathrm{lm}$ and different spline temporal basis functions. The quadratic spline temporal basis function with $\varepsilon_r = 3.2$ in (a) and $\varepsilon_r = 100$ in (b). The cubic spline temporal basis function with $\varepsilon_r = 3.2$ in (c) and $\varepsilon_r = 100$ in (d).}
\label{fig:EigSpline}
\end{figure}

As the Lagrange temporal basis functions do not result in a stable MOT-JVIE for high dielectric contrast, we now employ a different class of temporal basis functions: the spline temporal basis functions~\cite{Wang2007,VanTWout2013}. These temporal basis functions are $\mathcal{C}^{p-1}$-continuous, where $p$ is the temporal polynomial order. The linear spline temporal basis function is equal to the linear Lagrange temporal basis function and the stability performance is illustrated in Figures~\ref{fig:EigLin_epsr3_2} and~\ref{fig:EigLin_epsr100} in Section~\ref{sc:Lagrange}. We repeat the experiment of Section~\ref{sc:Lagrange} for the quadratic and cubic spline temporal basis function, shown in Figure~\ref{fig:Spline}, which are $C^1$-continuous and $C^2$-continuous, respectively. The interaction matrices are computed as explained in Section~\ref{sc:MatrixElements} for each of the temporal basis functions and the corresponding companion-matrix eigenvalue spectra are shown in Figure~\ref{fig:EigSpline}. The MOT-JVIE based on quadratic spline is stable in this numerical experiment independent of the relative permittivity. The MOT-JVIE based on cubic spline temporal basis functions is unstable. Unlike the MOT-JVIE based on Lagrange temporal basis functions, the MOT-JVIE based on cubic spline temporal basis functions remains unstable even when the dielectric contrast approaches zero.

\subsection{Discussion on continuity of the temporal basis function} \label{sc:EigDiscussion}
In the numerical experiments in the two previous sections we made the following observations regarding the stability of the MOT-JVIE: 1) the stability of Lagrange temporal basis functions improves as $\varepsilon_r$ approaches one; 2) the cubic spline temporal basis functions is unstable independent of $\varepsilon_r$; 3) the stability of the quadratic spline temporal basis functions is $\varepsilon_r$-independent. 

The first observation can be related to the discretized TDJVIE~\eqref{eq:MatrixEquation}. As $\varepsilon_r$ approaches one, the contribution of $\mathbf{C}_{m,m',n-n'}$ to the interaction matrices in~\eqref{eq:Zblocks} becomes smaller and we are left with the identity. Therefore, as $\varepsilon_r$ approaches one, the lower triangular matrix equation in~\eqref{eq:MatrixEquation} will become diagonally dominant and easier to solve. Consequently, the problem is related to the construction of $\mathbf{C}_{m,m',n-n'}$ in~\eqref{eq:Zblocks}, i.e. the discretization of
\begin{equation} \label{eq:curlcurlA}
    \nabla \times \mathbf{H}^s(\mathbf{r},t) = \nabla \times \nabla \times \iiint_{V_\varepsilon} \frac{\mathbf{J}_\varepsilon(\mathbf{r}',\tau)}{4 \pi R} \mathrm{d}V'
\end{equation}
in the TDJVIE~\eqref{eq:TDJVIE}. Thus the Lagrange temporal basis functions in combination with Dirac-delta temporal test functions and spatial piece-wise constant basis and test functions do not seem to be an appropriate choice for the temporal basis functions to discretize the term in Equation~\eqref{eq:curlcurlA}. As the MOT-scheme is constructed with Dirac-delta testing functions, there is continuity requirement on the temporal envelop of curl of the magnetic field strength induced by the contrast current density, i.e. $\nabla \times \mathbf{H}^s$ cannot contain Dirac-delta functions in time as these cannot be tested by the Dirac-delta functions we use to construct the MOT-scheme. Therefore, the induced magnetic field strength~\eqref{eq:MagFieldStrength} should be at least continuous, $\mathcal{C}^0$, in space-time to prevent these Dirac-deltas from occurring after the curl-operation. In the numerical example in Appendix~\ref{app:Lagrange}, we show a case to proof that the Lagrange temporal basis functions do not meet this $\mathcal{C}^0$-requirement. To obtain the $\mathcal{C}^0$-smoothness in the example we show, a smoother temporal basis function is required, like the quadratic spline. Thus, the required space-time smoothness of the magnetic field strength is at least of $\mathcal{C}^0$-continuity, which requires that temporal smoothness of the contrast current density is higher than $\mathcal{C}^0$-continuous. That is why the Lagrange temporal basis functions, which are all $\mathcal{C}^0$-continuous, results in a MOT-JVIE with $\varepsilon_r$-dependent stability. 

The second observation, i.e. the MOT-JVIE based on the cubic spline temporal basis function is always unstable, independent of $\varepsilon_r$, can be related to an incorrect temporal discretization of the $\varepsilon_r \mathbf{J}_\varepsilon$-term in Equation~\eqref{eq:TDJVIE}. As the scheme is unstable for all $\varepsilon_r$, we can study the case where $\varepsilon_r$ approaches one. In Appendix~\ref{app:CubicSpline}, we demonstrate analytically that a voxel with zero contrast can still support an exponentially increasing solution. One can say that the currently used Dirac-delta test procedure does not properly observe the $\varepsilon_r \mathbf{J}_\varepsilon$-term in Equation~\eqref{eq:TDJVIE} if expanded with cubic spline temporal basis function. On the contrary, in Appendix~\ref{app:CubicSpline} we prove that a voxel with zero contrast cannot support an exponentially increasing solution if the MOT-JVIE is based on the quadratic spline temporal basis function.

So, the right-hand side of~\eqref{eq:curlcurlA}, a term in Equation~\eqref{eq:TDJVIE}, requires a temporal basis function with a higher degree of continuity than $\mathcal{C}^0$ and the uniform Dirac-delta testing procedure does not properly observe the $\varepsilon_r \mathbf{J}_\varepsilon$-term in Equation~\eqref{eq:TDJVIE} when expanded with cubic splines. Combining these conclusions provides an explanation for our third observation, i.e. the stability of the quadratic spline temporal basis functions is $\varepsilon_r$-independent. 

\section{The MOT-JVIE with quadratic spline temporal basis function}~\label{sc:Results}
We now illustrate the capabilities of the MOT-JVIE solver based on quadratic spline temporal basis function through multiple numerical experiments. We use the scattering setup from Section~\ref{sc:ScatteringSetup}. To study the convergence of the solution for a finer discretization, we repeat numerical experiments for a range of values for $K$. Due to the piecewise constant spatial approximation for the contrast current density, the sample location has a significant influence on the convergence of the solution with respect to $K$. To minimize the effect of the sample location on the convergence study, we take the observation point always at the center of a voxel for all $K$. The reason behind using the center is that if we take a voxel with center $\mathbf{r}_m$ and decrease the dimensions of the respective voxel, the contrast current density inside the voxel should converge to the solution $\mathbf{J}_\varepsilon(\mathbf{r}_m)$ which solves~\eqref{eq:TDJVIE}. Consequently, we sample the contrast current density at $\mathbf{r}=(0.025,0.075,0.025)$ in all experiments, as this sample point is the center of one of the voxels for $K = 4$, $K = 12$ and $K = 20$.

\subsection{Literature Validation} \label{sc:Validation}
\begin{figure}[t]
\centering
\begin{subfigure}[b]{0.45\textwidth}
    \centerline{\includegraphics[width=\textwidth,draft=false]{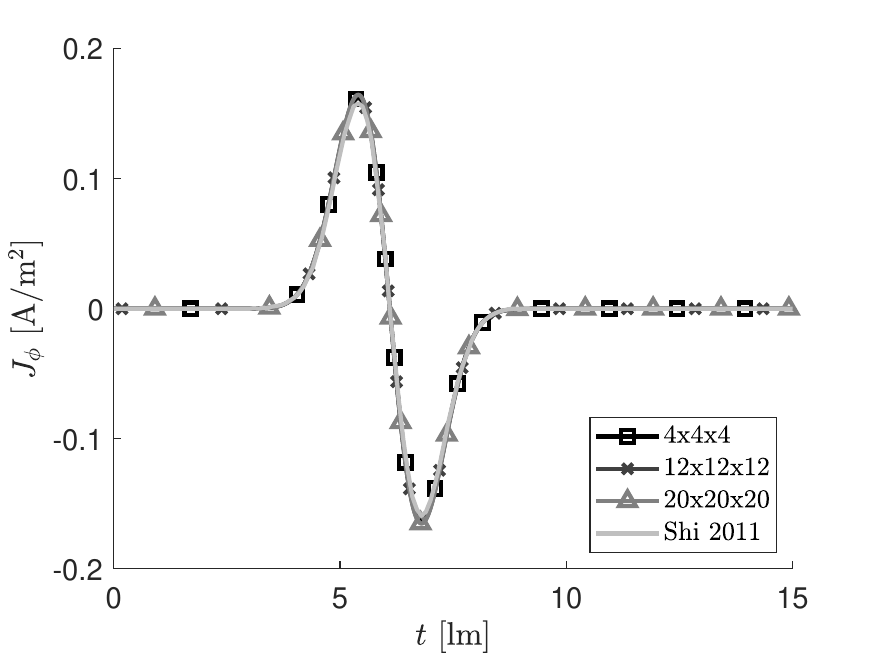}}
    \caption{}
    \label{fig:ShivsMOTJVIE}
\end{subfigure}
\begin{subfigure}[b]{0.45\textwidth}
    \centerline{\includegraphics[width=\textwidth,draft=false]{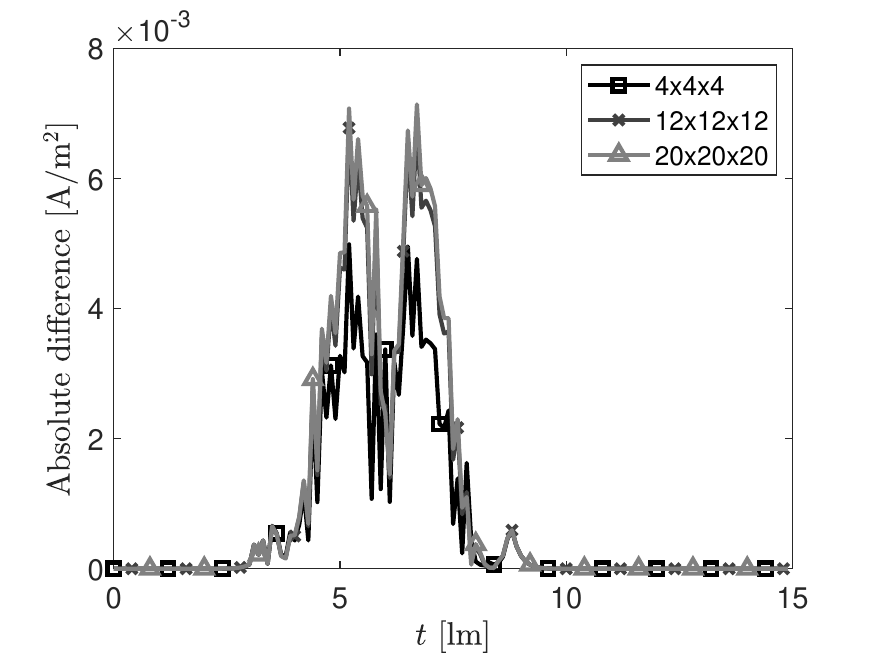}}
    \caption{}
    \label{fig:ShiValidationAbsDiff}
\end{subfigure}
\caption{(a) A comparison of the MOT-JVIE contrast current density solution excited by the Gaussian plane wave for a range of discretizations with the MOD-JVIE solution presented by Shi et al.~\cite{Shi2011dispersive}. Here, the contrast current density is sampled inside the $0.2^3$ dielectric cube with $\varepsilon_r = 3.2$ and can be expressed as $J_\phi(t) = \hat{\boldsymbol{\phi}} \cdot \mathbf{J}([0.025, 0.075, 0.025],t)$. (b) The absolute difference between the MOD-JVIE solution and the MOT-JVIE solutions shown in (a).}
\label{fig:ShiValidation}
\end{figure}

To study the validity of the MOT-JVIE implementation, we compare the MOT-JVIE solution to the Marching-on-in-Degree contrast current density volume integral equation (MOD-JVIE) solution presented by Shi et al. in~\cite{Shi2011dispersive}.
Here, the dielectric cube has a relative permittivity of $\varepsilon_r = 3.2$ and the excitation parameters are $w = 4~\mathrm{lm}$ and $t_0 = 6.1~\mathrm{lm}$. The induced contrast current density is sampled in the $\hat{\boldsymbol{\phi}}$-direction, i.e. the unit vector at the observation point $\mathbf{r}=(x,y,z)$ defined as $\hat{\boldsymbol{\phi}} = (-y,x,0)/\sqrt{x^2+y^2}$. The MOT-JVIE contrast current density solutions for an increasing number of voxels per Cartesian direction, $K$, and decreasing time step size, i.e. $\Delta t = 0.2/K ~\mathrm{lm}$, are shown in Figure~\ref{fig:ShivsMOTJVIE}, together with MOD-JVIE solution obtained from~\cite{Shi2011dispersive} using a plot digitizer~\cite{Rohatgi2021}. The absolute difference between the MOD-JVIE solution and the MOT-JVIE solution is given in Figure~\ref{fig:ShiValidationAbsDiff}.

The convergence of the MOT-JVIE solution for finer discretizations is observed in Figure~\ref{fig:ShiValidationAbsDiff} towards a different solution than the MOD-JVIE reference. Although the MOT-JVIE solution in Figure~\ref{fig:ShiValidation} is similar to the MOD-JVIE solution, a small difference is observed at the peaks. A difference in the MOT-JVIE and MOD-JVIE contrast current density solutions is to be expected as both spatial and temporal discretizations are different. Furthermore, the cube in~\cite{Shi2011dispersive} is discretized using only $K=4$ voxels and we expect that the MOD-JVIE solution in~\cite{Shi2011dispersive} has not yet fully converged. We conclude that the results of both solvers coincide up to the inaccuracies of their respective spatial discretizations, which establishes further confidence in the MOT-JVIE implementation.

\subsection{Frequency-domain validation} \label{sc:Accuracy}
\begin{figure}[t]
\centering
\begin{subfigure}[b]{0.32\textwidth}
    \centerline{\includegraphics[width=\textwidth,draft=false]{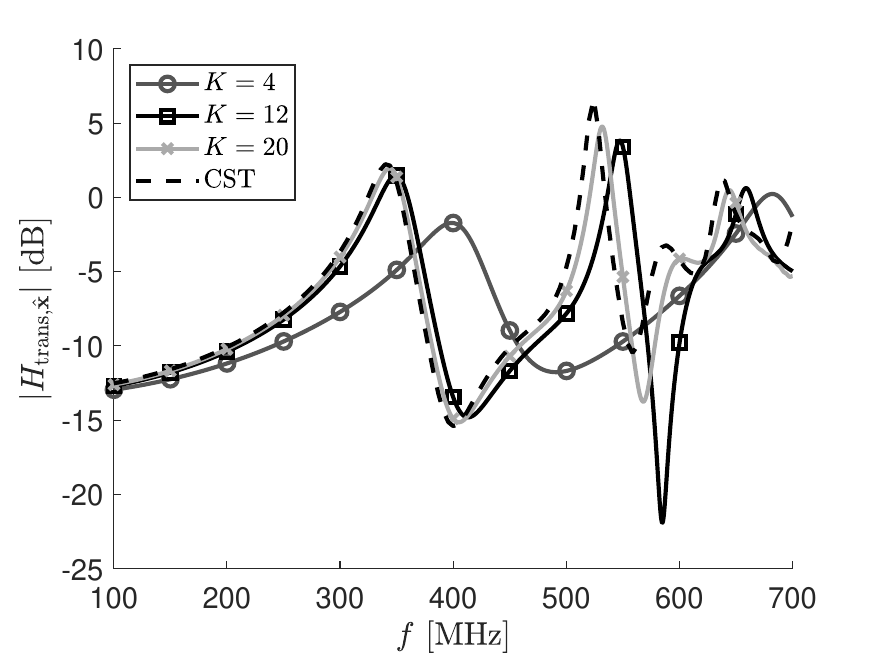}}
    \caption{}
    \label{fig:ExMOTJVIEvsCST}
\end{subfigure}
\begin{subfigure}[b]{0.32\textwidth}
    \centerline{\includegraphics[width=\textwidth,draft=false]{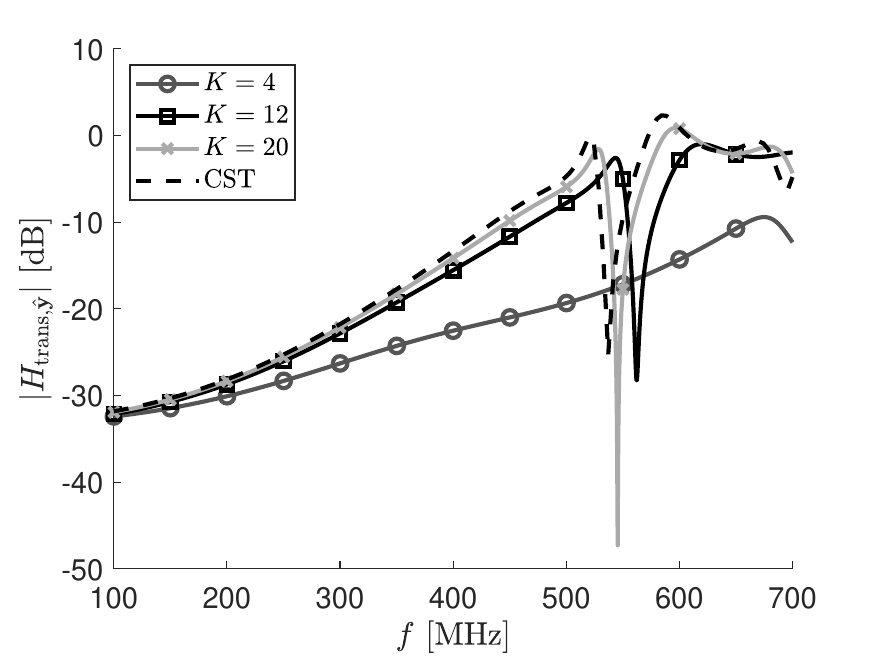}}
    \caption{}
    \label{fig:EyMOTJVIEvsCST}
\end{subfigure}
\begin{subfigure}[b]{0.32\textwidth}
    \centerline{\includegraphics[width=\textwidth,draft=false]{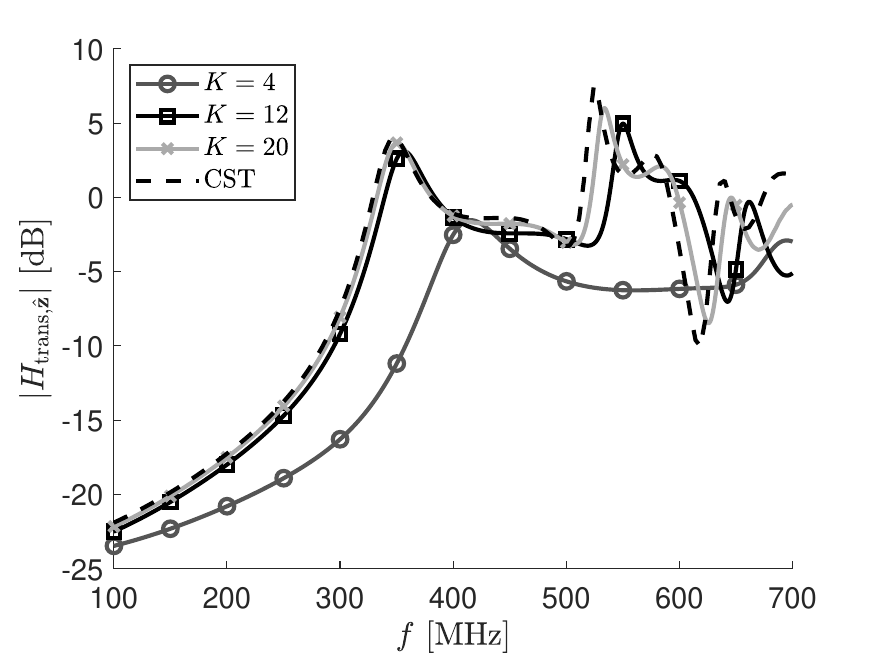}}
    \caption{}
    \label{fig:EzMOTJVIEvsCST}
\end{subfigure}
\caption{The magnitude of the frequency response of the electric field in the three Cartesian directions, i.e. (a) $|H_{\mathrm{trans},\hat{\mathbf{x}}}|$, (b) $|H_{\mathrm{trans},\hat{\mathbf{y}}}|$ and (c) $|H_{\mathrm{trans},\hat{\mathbf{z}}}|$, at $\mathbf{r}_m = (0.025, 0.075, 0.025)$ inside the dielectric cube with $\varepsilon_r = 12$, for a range of spatial discretizations $K$ and $\Delta t = 0.04~\mathrm{lm}$.}
\label{fig:MOTJVIEvsCST}
\end{figure}
We compare the frequency magnitude response based on the MOT-JVIE solution for increasing $K$ to the frequency magnitude response computed with CST Studio Suite 2023~\cite{CST2023} using its combined-field integral equation frequency-domain solver. To compute the frequency magnitude response in the scatterer, we employ the frequency-domain counterpart of~\eqref{eq:J_basis}, i.e.
\begin{equation}
    \mathbf{e}(\mathbf{r}_m,f) = \frac{1}{2 \pi \mathrm{j} f (\epsilon(\mathbf{r}_m) - \epsilon_0)} \mathbf{j}_\varepsilon(\mathbf{r}_m,f) ,
\end{equation}
where $\mathbf{j}_\varepsilon$ is the Fourier transform of $\mathbf{J}_\varepsilon$, which, using the Fourier transform of the quadratic spline temporal basis function in~\eqref{eq:QuadSpline}, is defined as
\begin{equation}
    \mathbf{j}_\varepsilon(\mathbf{r}_m,f) = \Delta t \, \mathrm{sinc}^3\left( \pi \Delta t f \right) \sum_{n=1}^{N} \mathbf{J}_{m,n}
    e^{- \mathrm{j} 2 \pi f n \Delta t }.
\end{equation}
The frequency magnitude response in the Cartesian direction $\hat{\mathbf{a}}$ is then equal to the transfer function $|H_{\mathrm{trans},\hat{\mathbf{a}}}|$, which is defined as
\begin{equation} \label{eq:FreqMagResponse}
|H_{\mathrm{trans},\hat{\mathbf{a}}}(\mathbf{r}_m, f)| = \frac{|\hat{\mathbf{a}} \cdot \mathbf{J}_\varepsilon(\mathbf{r}_m,f)|}{|e^i(f)|} ,
\end{equation}
where $|e_{\mathrm{i}}|$ is the magnitude of the Fourier transform of $\mathbf{E}^i \cdot \hat{\mathbf{x}}$, which is
\begin{equation}
    |e^i(f)| = \exp{\left( - \left( \frac{w}{4} \pi f  \right)^2  \right)}.
\end{equation}
Figure~\ref{fig:MOTJVIEvsCST} shows $|H_{\mathrm{trans},\hat{\mathbf{a}}}|$ for each of the Cartesian directions for the same range of $K$. The contrast current densities used to compute these values were generated with the following simulation settings: $\varepsilon_r = 12$, $w = 2~\mathrm{lm}$, $t_0 = 3.42~\mathrm{lm}$, $\Delta t = 0.04~\mathrm{lm}$, $N = 1500$. 

Two observations can be made from Figure~\ref{fig:MOTJVIEvsCST}: 1) independent of $K$ the accuracy of the MOT-JVIE solution deteriorates with increasing frequency and 2) the solution accuracy for a given frequency increases when increasing K. As we expect the accuracy to be proportional to the number of voxels per wavelength, both observations are in line with our expectations. Figure~\ref{fig:MOTJVIEvsCST} also illustrates a drawback of using piecewise-constant spatial basis functions, i.e. relatively slow convergence when increasing the number of voxels per wavelength. 

A similar experiment for analyzing the effect of a finer temporal discretization showed that the time step size is not the limiting factor regarding accuracy. The time step should therefore be chosen in accordance with the maximum frequency of the incident electromagnetic field.

\subsection{Long time sequence} \label{sc:FinerDiscretizationsAndStability}
\begin{figure}[t]
\centering
\begin{subfigure}[b]{0.32\textwidth}
    \centerline{\includegraphics[width=\textwidth,draft=false]{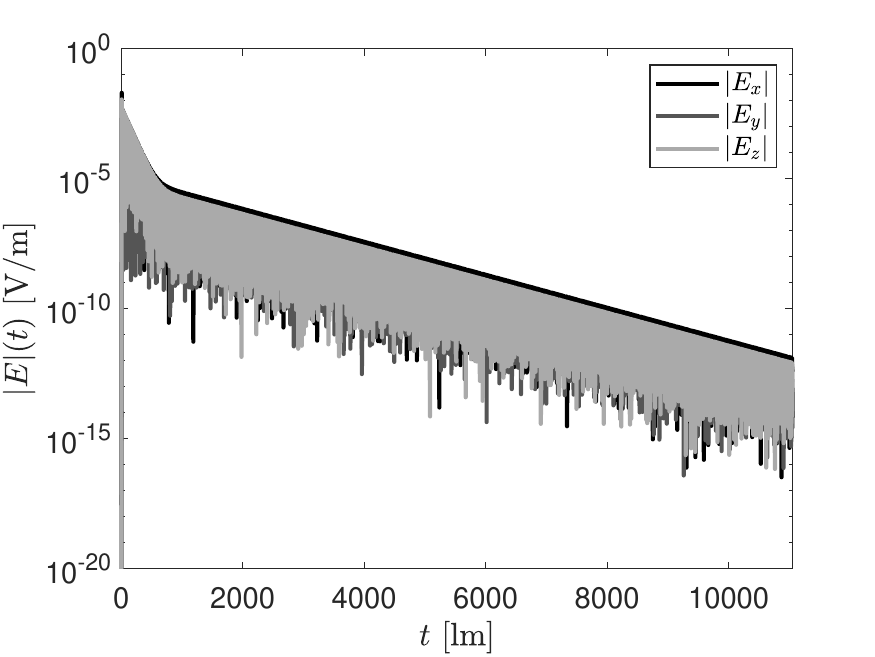}}
    \caption{}
    \label{fig:Epsr100_0-11050lm}
\end{subfigure}
\begin{subfigure}[b]{0.32\textwidth}
    \centerline{\includegraphics[width=\textwidth,draft=false]{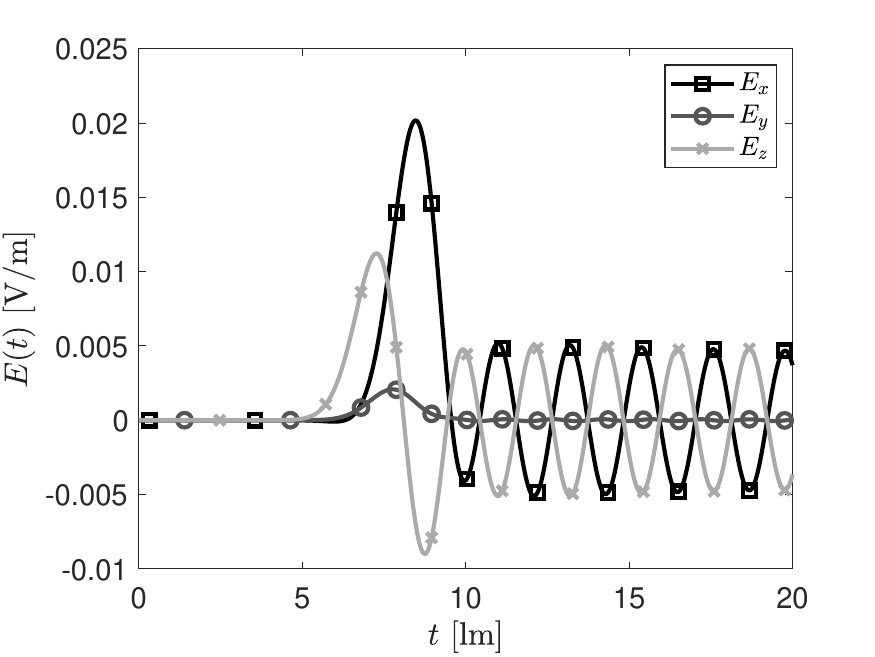}}
    \caption{}
    \label{fig:Epsr100_0-20lm}
\end{subfigure}
\begin{subfigure}[b]{0.32\textwidth}
    \centerline{\includegraphics[width=\textwidth,draft=false]{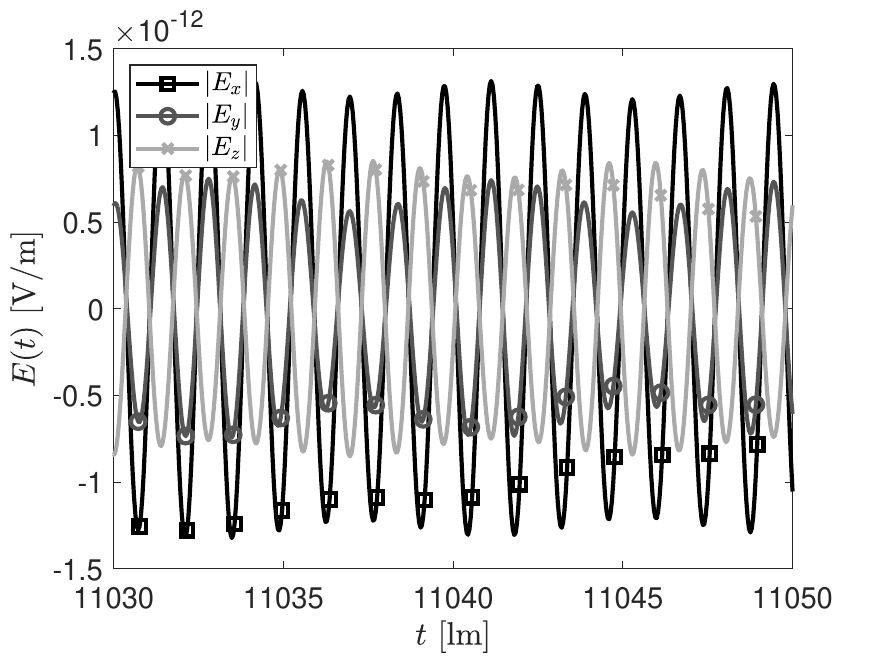}}
    \caption{}
    \label{fig:Epsr100_11030-11050lm}
\end{subfigure}
\caption{(a) Electric field magnitude, $|E|$, $(x,y,z)$-components at $\mathbf{r} = (0.025, 0.075, 0.025)$ inside the dielectric cube with $\varepsilon_r = 100$. The result is computed with the MOT-JVIE, where the cube was discretized with $K = 20$ and $\Delta t = 0.04~\mathrm{lm}$. (b) Zoom-in on the first $20~\mathrm{lm}$ of the solution shown in (a). (c) Zoom-in on the last $20~\mathrm{lm}$ of the solution shown in (a).}
\label{fig:Epsr100}
\end{figure}

In Section~\ref{sc:Spline} it was shown that the companion-matrix eigenvalues of the MOT-JVIE based on quadratic spline temporal basis functions for the dielectric cube with $\varepsilon_r = 3.2$ and $\varepsilon_r = 100$, discretized with $K = 6$ and $\Delta t = 0.2/K~\mathrm{lm}$, remain within the unit circle, even when increasing the dielectric contrast. 
As the companion-matrix stability analysis requires the eigenvalues of the companion matrix of dimension $K^3(\ell-1)$, this analysis is not feasible for $K>6$. Therefore, we choose to further demonstrate the MOT-JVIE performance for a finer discretization, i.e. $K = 20$ and $\Delta t = 0.04~\mathrm{lm}$, for the same dielectric cube with $\varepsilon_r = 100$, by computing the MOT-JVIE contrast current density for $276,480$ discrete time steps. For these spatial-temporal step sizes we expect the MOT-JVIE solution to be accurate only at low frequencies, yet it does illustrate the long-term behavior. The cube is excited by the Gaussian plane wave in~\eqref{eq:GaussPlaneWave} with $w = 5~\mathrm{lm}$ and $t_0 = 7.8~\mathrm{lm}$. The magnitude of the Cartesian components of the electric field at $\mathbf{r}=(0.025,0.075,0.025)$, resulting from the MOT-JVIE contrast current density, are shown on a logarithmic vertical scale in Figure~\ref{fig:Epsr100_0-11050lm}.

The magnitude of each Cartesian component of the electric field  decreases exponentially over time. The exponential decrease is steeper in the first $700~\mathrm{lm}$. A zoom-in on the time-axis, presented in Figures~\ref{fig:Epsr100_0-20lm} and~\ref{fig:Epsr100_11030-11050lm}, shows that lower frequencies are dominant in the first part of the signal and higher frequencies are dominant in the remaining part. This difference in frequency content explains the difference in exponential decrease, as there are multiple modes with different Q-factors and resonance frequencies. The most important observation to make from Figure~\ref{fig:Epsr100_0-11050lm} is that the electric field inside the dielectric cube decreases exponentially and no non-physical solutions are observed inside the high-dielectric-contrast scatterer.

\subsection{Inhomogeneous cube}
\begin{figure}[t]
    \centering
    \includegraphics[width=0.4\columnwidth]{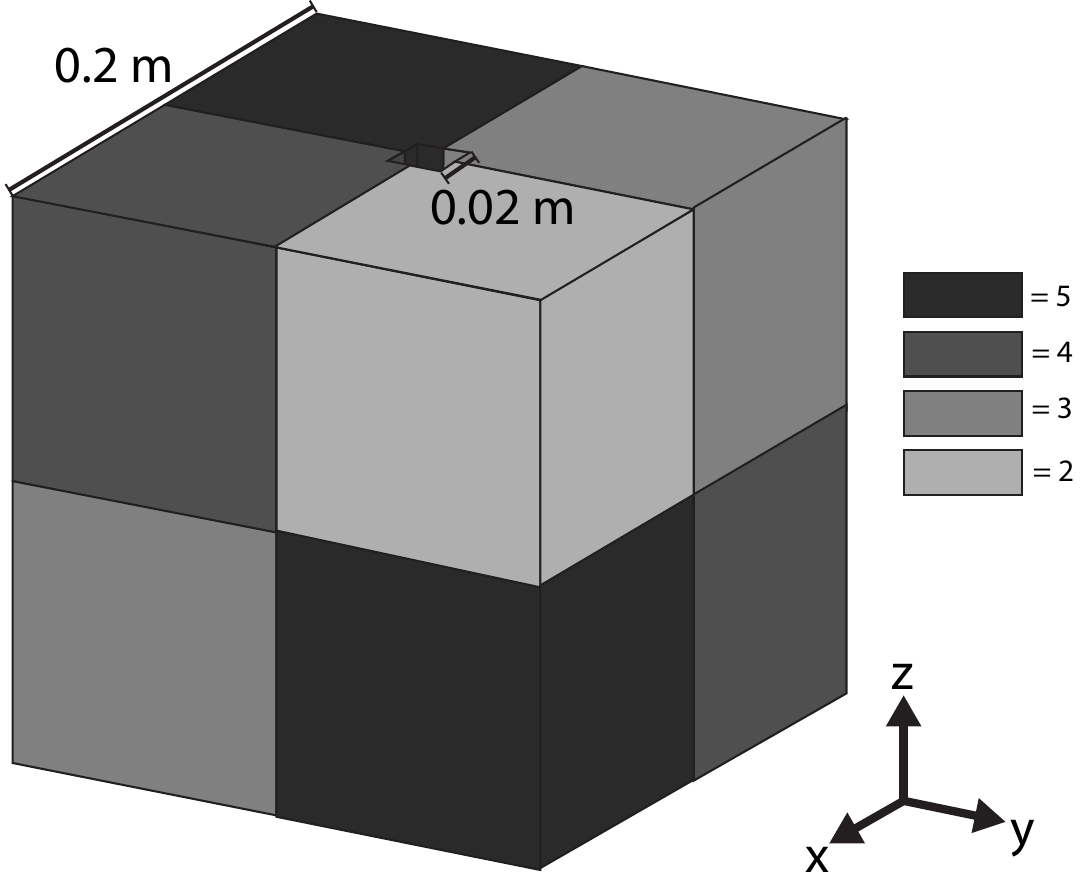}
    \caption{Inhomogeneous dielectric cube with edge length $0.2~\mathrm{m}$ with a square hole with edge length $0.02~\mathrm{m}$ from top to bottom. The cube is divided into eight equal cubes with permittivity values $\varepsilon_r = (2,3,4,5)$. The top half of the cubes is the same as the bottom half but rotated $180^\circ$ around the z-axis.}
    \label{fig:Inhomogeneous_cube}
\end{figure}

\begin{figure}[t]
\centering
\begin{subfigure}[b]{0.32\textwidth}
    \centerline{\includegraphics[width=\textwidth,draft=false]{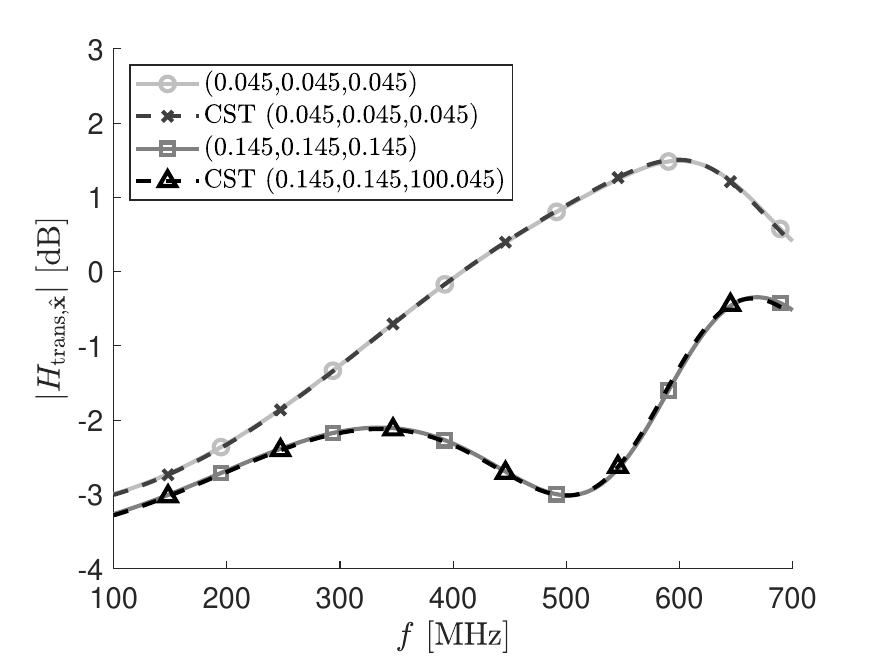}}
    \caption{}
    \label{fig:Inhomogeneous_ExEivsf_epsr2}
\end{subfigure}
\begin{subfigure}[b]{0.32\textwidth}
    \centerline{\includegraphics[width=\textwidth,draft=false]{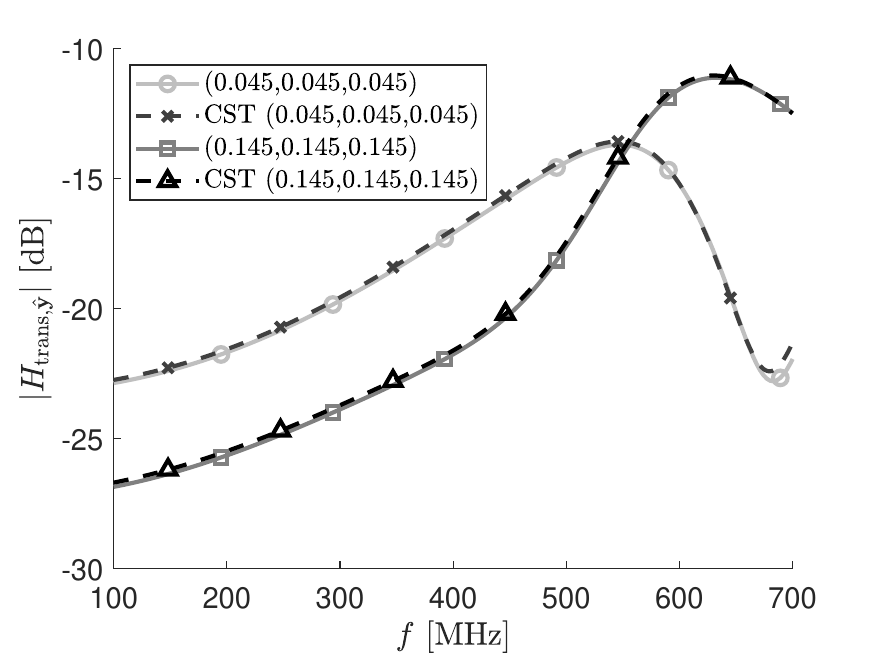}}
    \caption{}
    \label{fig:Inhomogeneous_EyEivsf_epsr2}
\end{subfigure}
\begin{subfigure}[b]{0.32\textwidth}
    \centerline{\includegraphics[width=\textwidth,draft=false]{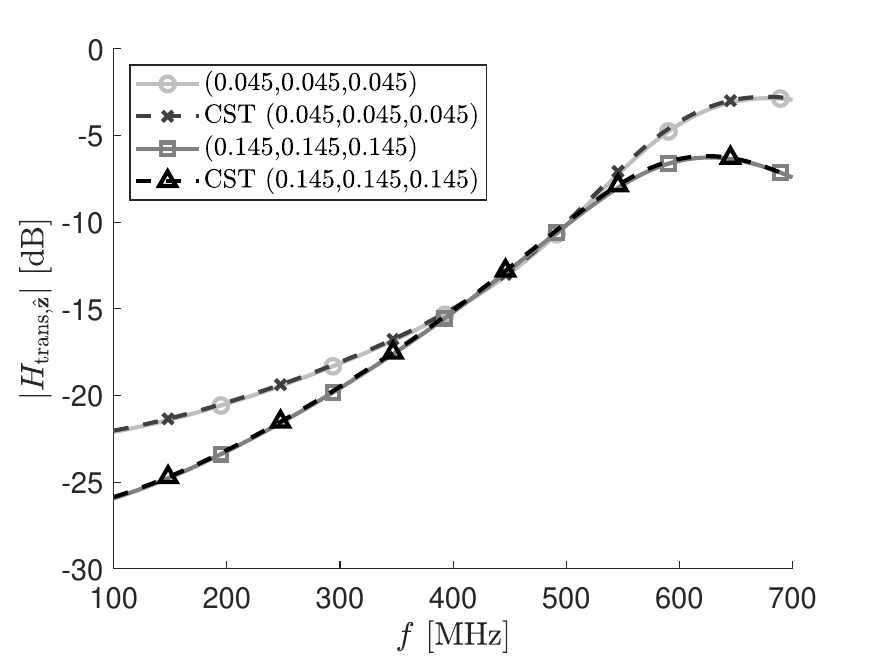}}
    \caption{}
    \label{fig:Inhomogeneous_EzEivsf_epsr2}
\end{subfigure}
\\
\begin{subfigure}[b]{0.32\textwidth}
    \centerline{\includegraphics[width=\textwidth,draft=false]{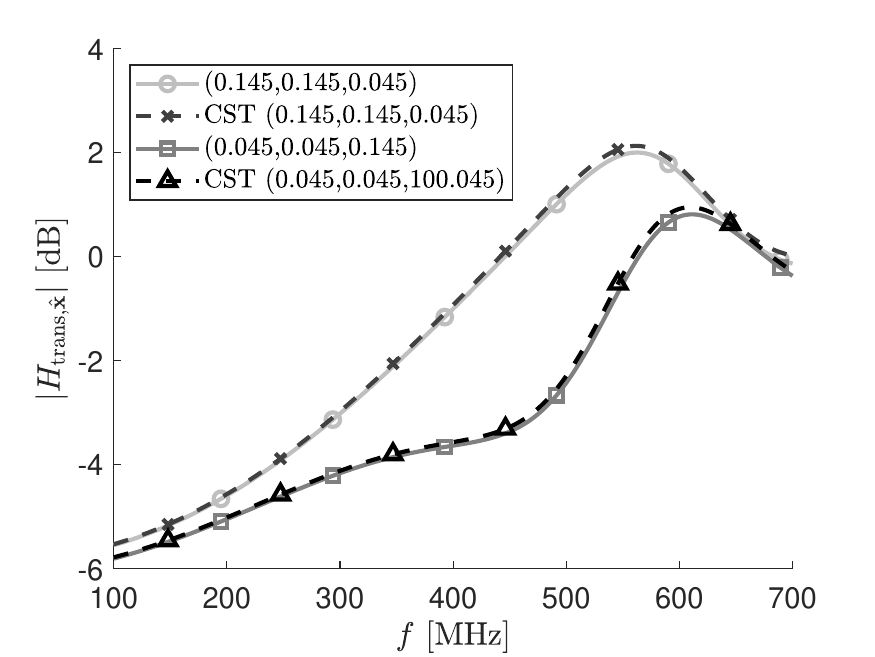}}
    \caption{}
    \label{fig:Inhomogeneous_ExEivsf_epsr5}
\end{subfigure}
\begin{subfigure}[b]{0.32\textwidth}
    \centerline{\includegraphics[width=\textwidth,draft=false]{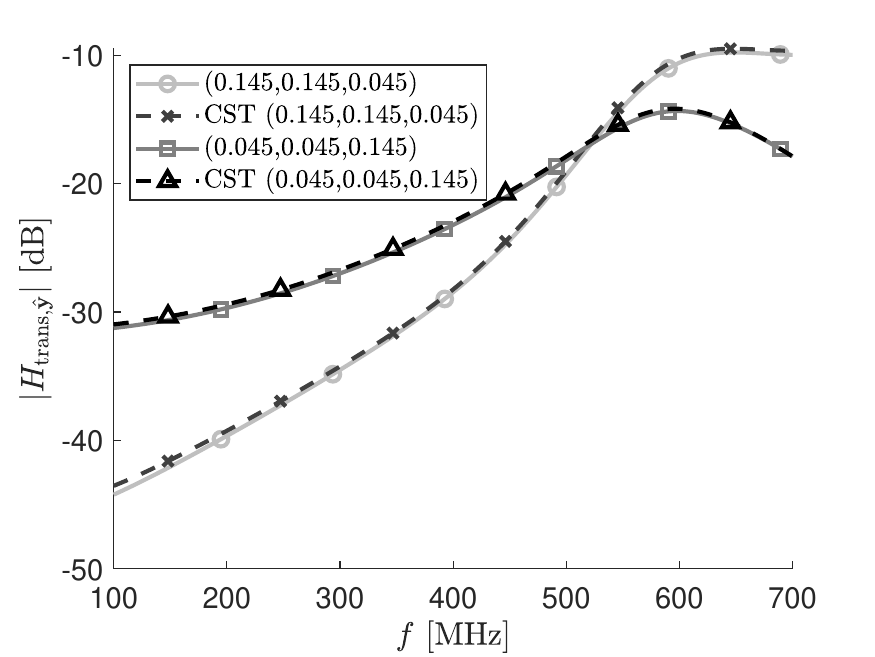}}
    \caption{}
    \label{fig:Inhomogeneous_EyEivsf_epsr5}
\end{subfigure}
\begin{subfigure}[b]{0.32\textwidth}
    \centerline{\includegraphics[width=\textwidth,draft=false]{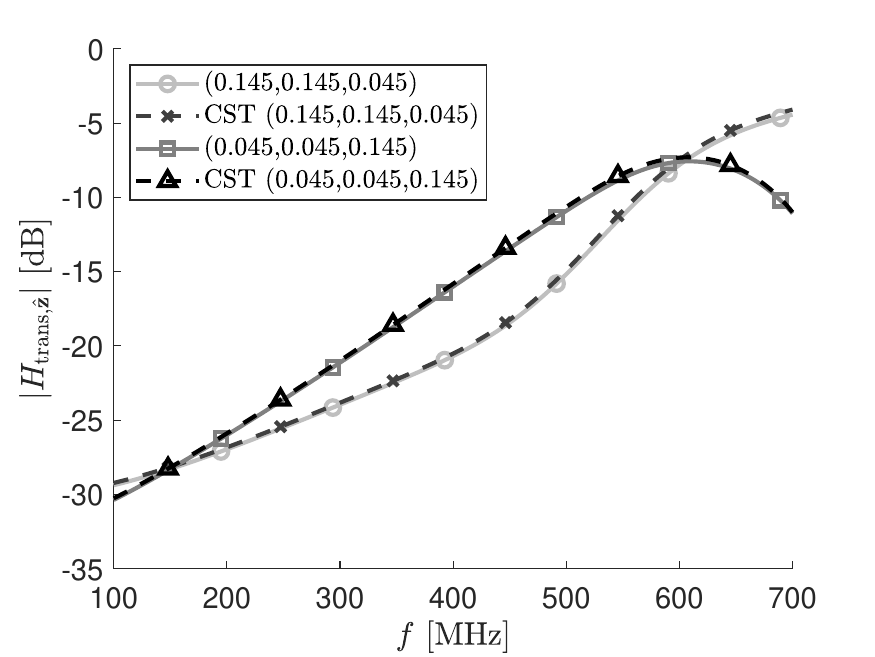}}
    \caption{}
    \label{fig:Inhomogeneous_EzEivsf_epsr5}
\end{subfigure}
\caption{The magnitude of the frequency response of the electric field in the three Cartesian directions, i.e. (a,d) $|H_{\mathrm{trans},\hat{\mathbf{x}}}|$, (b,e) $|H_{\mathrm{trans},\hat{\mathbf{y}}}|$ and (c,f) $|H_{\mathrm{trans},\hat{\mathbf{z}}}|$, at different locations inside the inhomogeneous dielectric cube shown in Figure~\ref{fig:Inhomogeneous_cube} discretized with $K=20$ and $\Delta t = 0.04~\mathrm{lm}$.}
\label{fig:InhomogeneouExyzEivsf}
\end{figure}

As the MOT-JVIE is based on the TDJVIE, it is also suitable to simulate scattering from an inhomogeneous scatterer. To illustrate, we re-use the interactions matrices computed in Section~\ref{sc:Accuracy} for $K=20$ and $\Delta t = 0.04~\mathrm{lm}$ and change the permittivity distribution of the cube to the inhomogeneous cube shown in Figure~\ref{fig:Inhomogeneous_cube}. In Figure~\ref{fig:InhomogeneouExyzEivsf} we show the frequency magnitude response based on the MOT-JVIE solution and the frequency magnitude response computed with the combined-field integral equation frequency-domain solver of CST Studio Suite 2023~\cite{CST2023} at four different locations in the inhomogeneous cube. As was also observed in Section~\ref{sc:Accuracy}, the accuracy is proportional to the number of voxels per wavelength and therefore the MOT-JVIE solution deteriorates as frequency increases.

\section{Conclusion} \label{sc:Conclusion}
The discretization of the time domain contrast current density volume integral equation with piece-wise constant spatial basis and test functions defined on voxels and Dirac-delta temporal test functions and piecewise polynomial temporal basis functions was presented. This discretization results in a causal implicit marching-on-in-time scheme to which we refer as the MOT-JVIE. A companion matrix stability analysis of the MOT-JVIE based on Lagrange and spline temporal basis functions indicated that, for a fixed spatial and temporal step size, the MOT-JVIE solver's stability is independent of scatterer's dielectric contrast for quadratic spline temporal basis functions. Whereas, Lagrange and cubic spline are unstable for high contrast. We related this stability performance to the expansion and testing procedure in time. The capabilities of the MOT-JVIE based on quadratic spline temporal basis functions were illustrated by: comparing the MOT-JVIE solution to time domain results from literature, which overlap up to the inaccuracies of the respective solvers, and frequency domain results from a commercial combined field integral equation solver, where similar trends in the solution are observed. Additionally, a long time sequence for a high dielectric contrast scatterer discretized with 24,000 spatial unknowns was simulated and no non-physical behavior was observed over the $276,480$ discrete time steps. 

\appendixx{Magnetic field strength $\mathcal{C}^0$-continuity requirement} \label{app:Lagrange}

\begin{figure}[t]
    \centering
    \begin{subfigure}[b]{0.32\textwidth}
        \centerline{\includegraphics[width=\textwidth,draft=false]{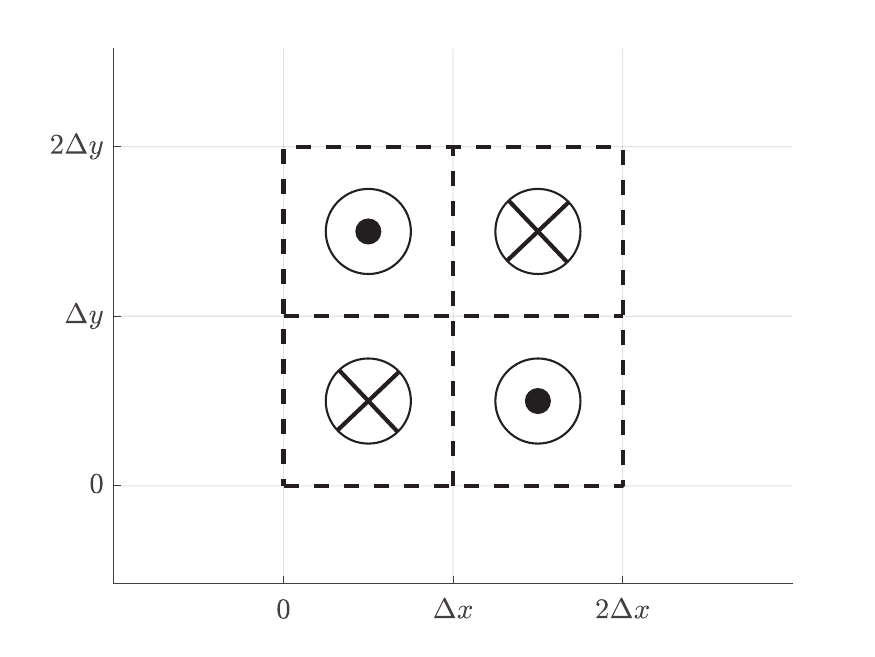}}
        \caption{}
        \label{fig:curlw}
    \end{subfigure}
    \begin{subfigure}[b]{0.32\textwidth}
    \centerline{\includegraphics[width=\textwidth,draft=false]{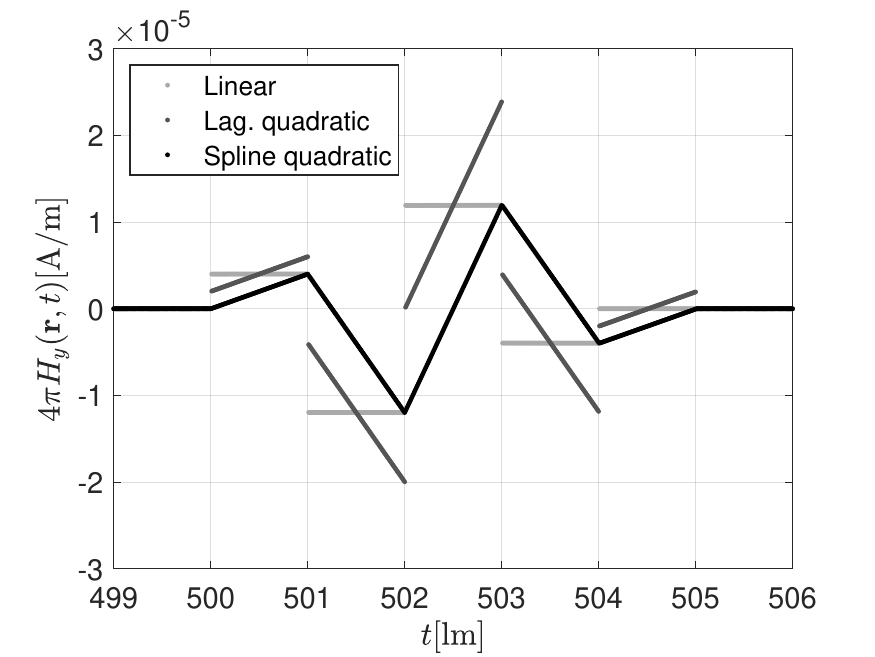}}
    \caption{}
    \label{fig:HTime}
    \end{subfigure}
    \begin{subfigure}[b]{0.32\textwidth}
    \centerline{\includegraphics[width=\textwidth,draft=false]{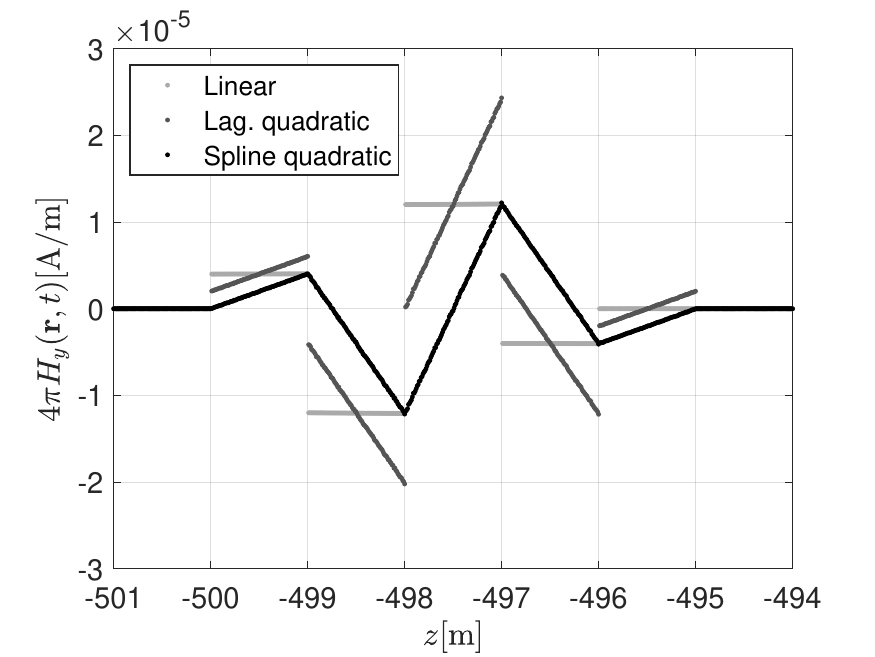}}
    \caption{}
    \label{fig:HSpace}
    \end{subfigure}
    \caption{(a) A set of $2\times2$ adjacent voxels defined in the (x,y)-plane represented by the black dashed lines with a piece-wise constant unit contrast current density defined in each voxel. The discretization settings are $\Delta x = \Delta y = \Delta z =  1 ~\mathrm{m}$ and $\Delta t = 1~\mathrm{lm}$. (b) The magnetic field strength in $\hat{\mathbf{y}}$-direction, $H_y$, sampled at $\mathbf{r} = (-500,0,0)$ and time instances $t = 499,499.01,499.02,\ldots,506$ induced by the contrast current distribution shown in (a). (c) The magnetic field strength in $\hat{\mathbf{y}}$-direction, $H_y$, sampled at $t = 500~\mathrm{lm}$ and space instances $\mathbf{r} = (z,0,0)$ with $z = -494,-494.01,-494.02,\ldots,-501$ induced by the contrast current distribution shown in (a). }
    \label{fig:Helmholtz_curlw}
\end{figure}

We compute the magnetic field strength, $\mathbf{H}^s$, produced by the current distribution shown in Figure~\ref{fig:curlw} with a linear and quadratic Lagrange and a quadratic spline time envelope. The dominant $\hat{\mathbf{y}}$-component, $H_y$, is shown in Figure~\ref{fig:HTime} and~\ref{fig:HSpace} as a function of time and space, respectively. The magnetic field strength of the Lagrange temporal basis functions are discontinuous in space-time and therefore do not meet the $\mathcal{C}^0$-requirement. Any higher order Lagrange temporal basis function would have the same discontinuity and is therefore not included in the numerical experiment. On the other hand, the smoother quadratic spline temporal basis function does meet the $\mathcal{C}^0$-requirement on the magnetic field strength. Any higher order spline would result in a smoother magnetic field strength, but the quadratic spline has the minimum continuity to obtain a continuous magnetic field strength in space-time. 

\appendixx{Dirac-delta test and spline temporal basis function} \label{app:CubicSpline}
To illustrate that the Dirac-delta test function in combination with the cubic spline temporal basis function leads to unstable MOT-scheme, let us look at discretization of a cube with a permittivity approaching one, i.e. $\varepsilon_m \rightarrow 1$ for $m = 1,\ldots,M$. In that case, there is no interaction between voxels nor between the different Cartesian components of the current density in a single voxel, i.e. only the identity-term in Equation~\eqref{eq:Zblocks} remains with $T(n\Delta t) = \frac{1}{6},\frac{2}{3},\frac{1}{6}$ for $n = 0,1,2$ and zero for all other integers $n$~\cite{VanTWout2013}. Therefore, the MOT-JVIE~\eqref{eq:MOT-JVIE} can be updated per voxel and per Cartesian direction and the update scheme simplifies to
\begin{equation}
    J^\alpha_{m',n} = - 4 J^\alpha_{m',n-1} -  J^\alpha_{m',n-2}.
\end{equation}
The companion-matrix eigenvalues belonging to this MOT-scheme are $\lambda_1 = -2-\sqrt{3}$ and $\lambda_2 =-2+\sqrt{3}$. As $|\lambda_1|>1$, the scheme admits an exponentially increasing solution when the TDJVIE is discretized with the cubic spline temporal basis function and the Dirac-delta test function. 

The same analysis can be performed for the Dirac-delta test function in combination with quadratic spline basis function. The MOT-scheme in that case simplifies to
\begin{equation}
    J^\alpha_{m',n} = - J^\alpha_{m',n-1}.
\end{equation}
The companion-matrix eigenvalue is then $\lambda = -1$. As $|\lambda| = 1$, the scheme does not admit exponentially increasing solutions when the TDJVIE is discretized with the quadratic spline temporal basis function and the Dirac-delta test function.

\bibliographystyle{ieeetr}
\bibliography{bibliography.bib}

\end{document}